\documentclass[showpacs, pra, aps, twocolumn,10pt]{revtex4}
\usepackage{mathrsfs}
\usepackage{array}
\usepackage{amsmath}
\usepackage{graphicx}
\usepackage{amstext}
\usepackage{bm}
\usepackage{amsfonts}
\usepackage{color}

\begin{document}
\title{Steady-state tripartite non-Gaussian entanglement and steering in output field from intracavity triple-photon parametric downconversion}
\author{Miaomiao Wei}
\affiliation{Department of Physics, Huazhong Normal University, Wuhan 430079, China}
\author{Huatang Tan}
\email{tht@mail.ccnu.edu.cn}
\affiliation{Department of Physics, Huazhong Normal University, Wuhan 430079, China}

\begin{abstract}
Nondegenerate triple-photon parametric downconversion (NTPD) is a potential source for unconditional tripartite non-Gaussian entangled states of continuous variables. Recent experiment has demonstrated strong third-order correlations among bright photon triplets via microwave NTPD in a superconducting cavity [Phys. Rev. X 10, 011011 (2020)]. Previous theoretic works have revealed that only short-time genuine tripartite non-Gaussian entanglement can be generated in NTPD even in the absence of dissipation. In this paper, we  investigate the properties of tripartite non-Gaussian entanglement and steering in the cavity output field by taking into account of the cavity dissipation. We first derive experimentally detectable criteria for fully inseparable and genuine tripartite non-Gaussian entanglement and steering. With the criteria, we then find that steady-state tripartite non-Gaussian entanglement and steering can be generated in the output field, although they merely exist in the short-time regime inside the cavity.  We also find that the initial cavity-field coherent states can obviously enhance the steady-state and transient tripartite entanglement and steering, in comparison to the case of initial vacuum states.
We finally show that the output tripartite non-Gaussian steerable correlations can be applied to the remote generation of negative Wigner-function quantum states by homodyne detection.  \end{abstract}
\maketitle

\section{Introduction}
Entanglement, a fundamental property within the domain of quantum mechanics, describes the inseparability inherent in composite quantum systems consisting of multiple constituent elements and is a vital resource in quantum information science \cite{Q1}. The concept of quantum steering traces its origin back to 1935, originally termed by Sch\"{o}dinger in his response to the well-known Einstein-Podolsky-Rosen paradox (EPR) \cite{en} to critique the nonlocal aspects of quantum mechanics. It has been verified that EPR steering is intermediate between
Bell nonlocality and entanglement \cite{enst} and useful in e.g. one-sided device-independent quantum cryptography \cite{Oneside}, subchannel discrimination
\cite{sub1,sub2}, and secure quantum teleportation \cite{secure}. Recent studies have further shown that Gaussian steering is a sufficient and necessary condition for remotely creating negative-Wigner nonclassicality on certain conditions\cite{WN1, WN2}. Steering has nowadays been realized in a variety of systems of discrete and continuous variables \cite{steer1,steer2}. 

Non-Gaussian entanglement of continuous variables is of paramount importance in various aspects of quantum science \cite{nG1, nG2, nG3, nG4}. Non-Gaussian entangled states feature diverse high-order moments of field quadrature operators, beyond second-order moments statistics in Gaussian states, resulting in its advantages in the aspects e.g. fundamental test of quantum mechanics such as loophole-free Bell test \cite{HD}, quantum error correction \cite{Error}, entanglement distillation \cite{Distillation}, and especially universal quantum computation \cite{Computation}. Non-Gaussian entangled states have also proven to be more efficient in quantum communication  \cite{appli1, distri2, distri3, telepo} and quantum sensing and metrology \cite{non5, metro}.
Over the past decades, the generation of non-Gaussian entangled states via photon-addition or -subtraction operation on Gaussian states has been extensively studied theoretically and experimentally \cite{pa1,pa2,ps1,ps2,ps3,ps4,ps5}, but this approach is probabilistic and the target states are conditioned on the detection results. Alternative way is to employ intrinsic nonlinearity of systems to achieve unconditional non-Gaussian states \cite{qe1,qe2,qe3,qe4,qe5,qe6}.

NTPD describes a nonlinear process in which a pump photon is downconverted into photon triplets of different frequencies and is considered as a potential source for  deterministically generating tripartite non-Gaussian highly entangled states directly \cite{three1,three2,three3,three4,three5,three6,three7,three8}. So far, NTPD process has been demonstrated in different three-order optical nonlinear mediums but with low rates of triple photon generation, which makes it difficult to certify quantum features \cite{OW1,OW2,OW3,Bench}. Very recently, microwave NTPD in a superconducting cavity has been achieved and strong third-order correlations among bright photon triplets has been demonstrated \cite{Sandbo}. This achievement immediately attracts much interesting in exploring the properties of tripartite non-Gaussian entanglement of continuous variables \cite{wil,three9,Tian,wei,Da}. It has been revealed that genuine tripartite non-Gaussian entanglement can be directly generated but it just appears in the short-time regime, even without the consideration of dissipation. In view that the NTPD process operates in the cavity in the experiment \cite{Sandbo}, quantum steering is stronger than entanglement and steady-state entanglement is more desirable,  a question naturally arises: Whether does the cavity output field exhibit steady-state tripartite non-Gaussian entanglement and even steering? As we know, the cavity output field is a continuum of frequency modes and indeed subject to realistic detection and various applications, with different behaviors from the intracavity field \cite{JP,YL}.



\begin{figure}[t]
\vspace{0cm}
\centerline{\hspace{1cm}\scalebox{0.27}{\includegraphics{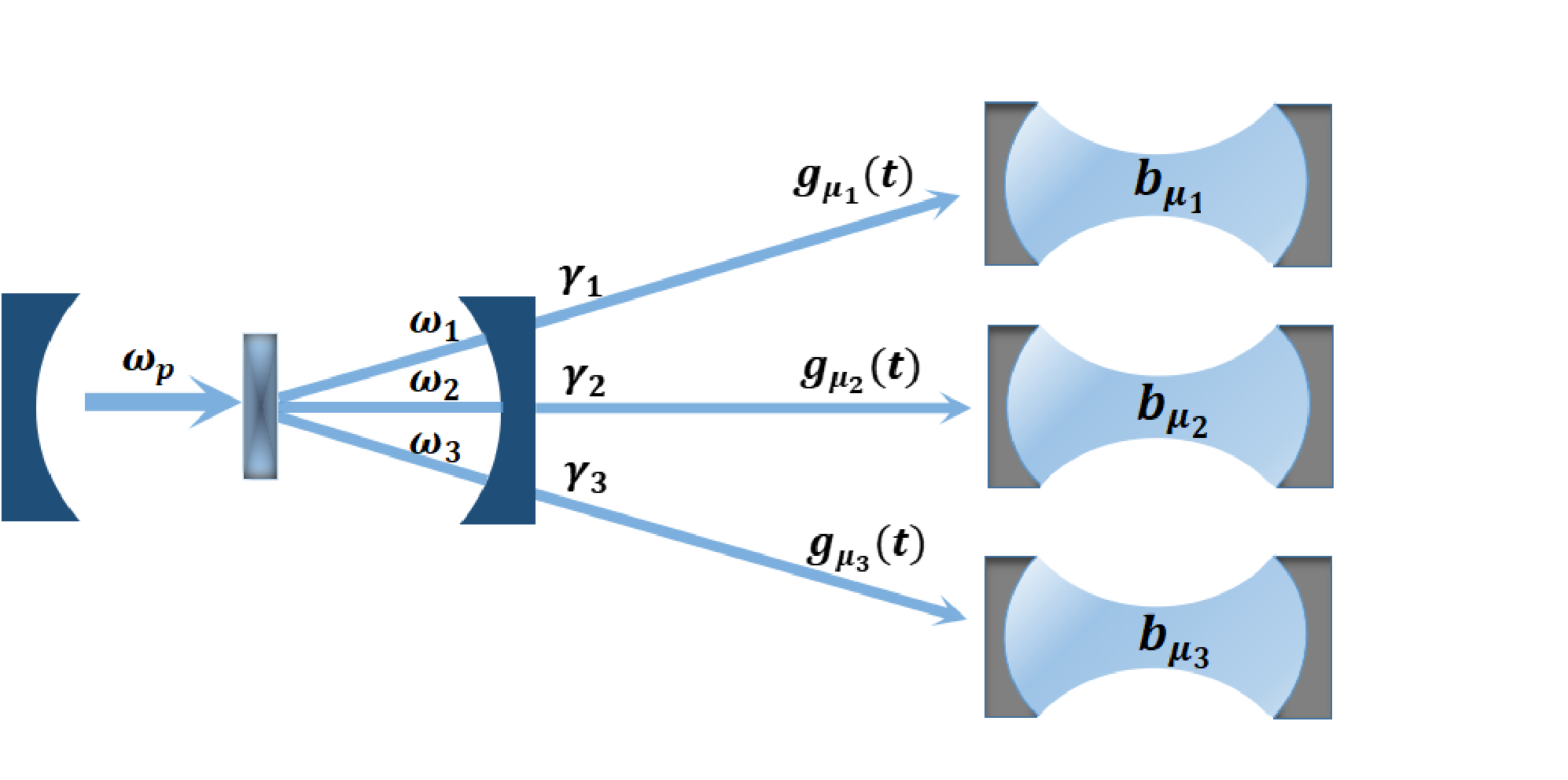}}}
\vspace{0cm}
\caption{Schematic diagram for intracavity NTPD in which a high-frequency pump photon (with frequency $\omega_p$) is downconverted into a triplet [(denoted by $\hat b_k(k=1,2,3)]$ in the cavity modes of frequencies $\omega_k$. The parameters $\gamma_k$ denote the dissipation rates of the cavity modes. Three virtual cavities, with modes denoted by $\hat b_{\mu_k}$, are employed to study the specific temporal modes $\mu_k(t)$ in the continuous cavity output fields $\hat b_k^{\rm{out}}(t)$ via the cascaded couplings to the (master) cavity modes $\hat b_k$, with the time-dependent coupling rates $g_{\mu_k}(t)$ dependent the modes $\mu_k(t)$. }
\label{fig1}
\end{figure}
In this paper, we intend to investigate in detail the properties of tripartite non-Gaussian entanglement and steering in the cavity output field in the system of intracavity NTPD.
To this end, we treat the output field with specific modes as virtual cavities connected to the NTPD cavity in a way of a quantum input-output (cascade) network \cite{Alexander,Alex}. We also derive experimentally detectable criteria for fully inseparable and genuine tripartite non-Gaussian entanglement and steering. With the criteria, we find that the steady-state tripartite non-Gaussian entanglement and steering can be generated in the output field, although they merely exist in the short-time regime inside the cavity. Moreover, the initial coherent cavity-field states can effectively enhance the output steady-state and intracavity transient entanglement and steering.
We also show that the output tripartite non-Gaussian steerable nonlocality can be used to remotely generate negative Wigner-function non-Gaussian states by homodyne detection. Our findings further unravel the novel non-Gaussian nonclassical characteristics in the nonlinear NTPD process.

This paper is arranged as follows. In Sec. II, the system is introduced and the master equation is given. In Sec. III, the criteria for fully inseparable and genuine tripartite non-Gaussian entanglement and steering are derived in detail. In Sec. VI, the numerical results are presented. In Sec. V, the summary is given.

\section{System}
In this paper, we consider an intracavity nondegenerate three-photon downconverion process, which can be described by the Hamiltonian ($\hbar=1$) \cite{Bench, Sandbo}
\begin{align}
\hat H_s&=\omega_0 \hat b_0^\dag \hat b_0+ \sum_{k=1}^3\omega_k \hat b_k^\dag\hat b_k+g_0(\hat b_0^\dag\hat b_1 \hat b_2 \hat b_3 \ {+}\hat b_0\hat b_1^{\dag}\hat b_2^{\dag}\hat b_3^{\dag}),
\label{eq11}
\end{align}
where $g_0$ represents the three-order nonlinear coupling constant, and the annihilation operators $\hat b_0$ and $\hat b_k~(k=1,2,3, \text{and similarly hereinafter})$ describe the
the pump and three down-converted modes, respectively. By choosing the frequencies $\omega_0=\omega_1+\omega_2+\omega_3$ and treating the pump classically (assuming it in a large-amplitude coherent state), the above interaction
Hamiltonian reduces to
\begin{align}
\hat H_s&=g(e^{i\theta}\hat b_1 \hat b_2 \hat b_3+e^{-i\theta}\hat b_1^{\dag}\hat b_2^{\dag}\hat b_3^{\dag}),
\label{eq1}
\end{align}
where $g=g_0|\beta_0|$ represents the NTPD interaction strength proportional to the pump amplitude $\beta_0\equiv|\beta_0|e^{i\theta}$. Here, we take $\theta=\pi/2$ for simplicity. Note that the phase $\theta$ can be cancelled via the local transformation $\hat b_j e^{-i\theta}\rightarrow \hat b_j$, which does not alter the tripartite correlations. Microscopically, it describes that the medium absorbs a high-frequency pump photon and then emits three low-frequency photons simultaneously into the cavity modes, during which strong non-Gaussian quantum correlations are therefore be established among the downconverted photons. The NTPD process has been demonstrated in optical nonlinear mediums \cite{Bench} and in a superconducting device \cite{Sandbo}.


For the cavity mode $\hat b_k$ coupled to external environment, one is interested in quantum properties of its output field which is indeed subject to detection and various realistic applications. The output field $\hat b_k^{\rm out}$ is related to the cavity mode $\hat b_k$ and input field $\hat b_k^{\rm in}$ via the input-output relation $\hat b_k^{\rm out}(t)=\sqrt{\gamma_k}\hat b_k(t)-\hat b_k^{\rm in}(t)$, where $\gamma_k$ denote the dissipation rate of the cavity mode and $\hat b_k^{\rm in}$ is the vacuum input. Since the cavity output field has continuous spectra, from which one can define a temporal mode $\mu_k(t)$ with the annihilation operator
\begin{align}
\hat b_{\mu_k}=\int \mu_k^*(t')\hat b_k^{\rm out}(t')dt',
\label{eq2}
\end{align}
which satisfies the commutation relation $[\hat b_{\mu_k}, \hat b_{\mu_k}^\dag]=1$, leading to $\int |\mu_k(t)|^2dt=1$. The mode $\hat b_{\mu _k}$ filtered from the output field $\hat b_k^{\rm out}(t)$ can be considered as a virtual cavity (filter) which is directionally driven by the output field, as shown in Fig.1, with the coupling strength between the output field and the virtual cavity \cite{Alexander}
\begin{align}
g_{\mu_k}(t)=-\frac{\mu_k^*(t)}{\sqrt{\int_0^{t} dt'|\mu_k(t')|^2}}.
\label{eq3}
\end{align}
In this description, the master equation for the whole cascaded system consisting of the (master) cavity mode $\hat b_k$ in the NTPD and the corresponding (slave) virtual cavity field $\hat b_{\mu_k}$ can be obtained as  \cite{input2}
\begin{equation}
\begin{split}
\frac{d\hat \rho}{dt}=-i[\hat H_s+\hat H_{ex},\hat \rho]+\sum_{k=1}^3\mathcal L_k[\hat J_k]\hat \rho,
\end{split}
\label{me}
\end{equation}
where the unidirectional-coupling resulted coherent exchange couplings
\begin{equation}
\begin{split}
\hat H_{ex}&=\frac{i}{2}\sum_{k=1}^3(\sqrt{\gamma_k}g_{\mu_k}^*\hat b_k^\dag\hat b_{\mu_k}-h.c.),
\end{split}
\end{equation}
and the collective decay $\mathcal L[\hat J_k]\hat \rho=\hat J_k\hat \rho\hat J_k^\dag-\frac{1}{2}(\hat J_k^\dag\hat J_k\hat \rho+\hat \rho\hat J_k^\dag\hat J_k)$, with the jump operators
\begin{align}
\hat J_k=\sqrt{\gamma_k}\hat b_k+g_{\mu_k}^*\hat b_{\mu_k},
\end{align}
describing the collective dissipation due to the couplings of the downconverted cavity and virtual cavity modes to the corresponding common vacuum reservoirs. With Eq.(\ref{me}), we can study quantum correlations in the intracavity and output fields. Here the six-mode master equation will be solved numerically by using quantum-jump Monte-Carlo approach to reduce  the Hilbert space  dimensions.
In this setting, the master equation (\ref{me}) can be unveiled by considering that the virtual cavity modes are monitored via continuous photon counting and the state of the whole system on one quantum trajectory can be described by the state vector \cite{MC3}
\begin{equation}
\begin{split}
d|\psi(t)\rangle&=\sum_{k=1}^{3}\Big[dN_k(t)\Big(\frac{\hat{J}_k}{\sqrt{\big\langle\hat{J}_k^\dag\hat{J}_k}\big\rangle}-\hat{1}\Big)
+dt\Big(\frac{\big\langle\hat{J}_k^{\dag}\hat{J}_k\big\rangle(t)}{2}\\&-\frac{\hat{J}_k^{\dag}\hat{J}_k}{2}-i\hat{H}\Big)\Big]|\psi(t)\rangle,
\end{split}
\end{equation}
where $\hat{H}=\hat H_s+\hat H_{ex}$ and the stochastic increment $dN_k(t)$ is either one or zero, representing (no) registration of photons of the detector. The density matrix $\hat \rho$ is obtained by performing ensemble average on different quantum trajectories. Here, unless otherwise stated, the ensemble average is done with two thousand trajectories. In addition, to solve the equation we consider the initial states of the downconverted modes to be vacuum or coherent states and the virtual cavity modes to be vacuum states.

\begin{figure}[t]
\vspace{0cm}
\centerline{\hspace{-0.25cm}\scalebox{0.225}{\includegraphics{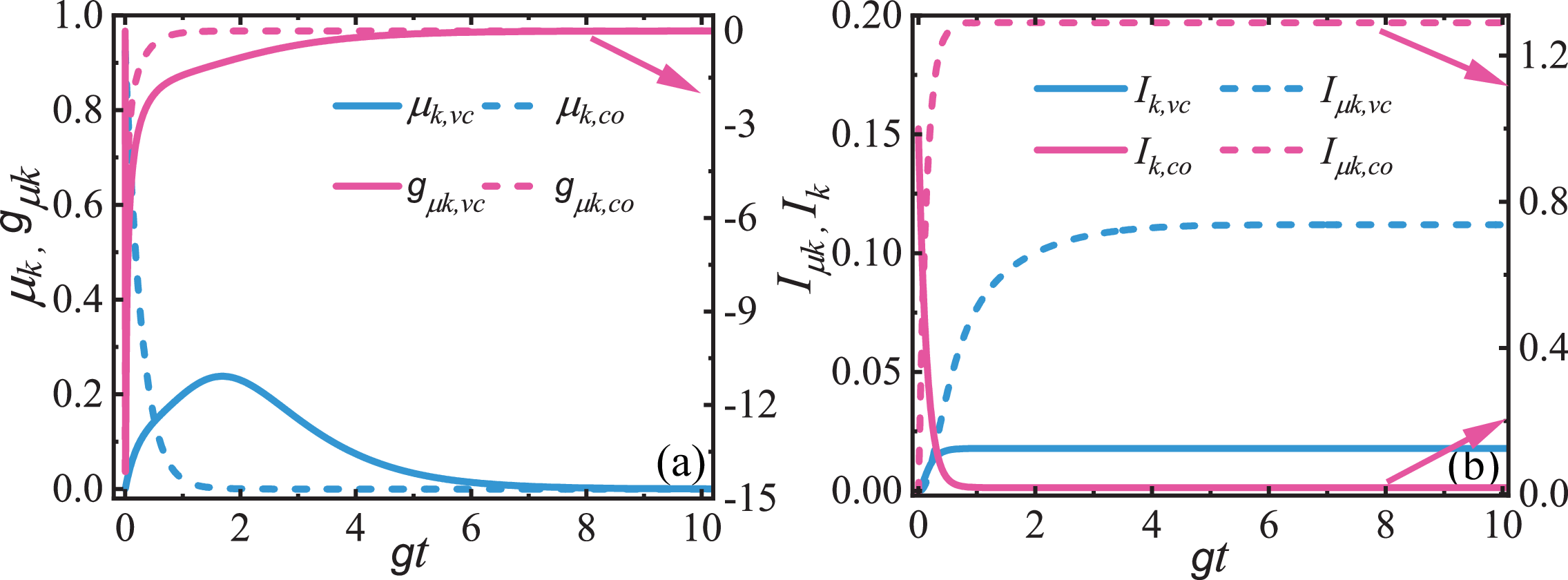}}}
\vspace{0cm}
\caption{(a) The most populated output modes $\mu_k(t)$ and the coupling $g_{\mu_k}$ for initial vacuum and coherent states of the intracavity modes $\hat b_k$, respectively.
(b) The mean photon numbers $I_k=\langle\hat b_k^\dag\hat b_k \rangle$ and $I_{\mu k}=\langle\hat b_{\mu_k}^\dag\hat b_{\mu_k} \rangle$ of the intracavity modes $\hat b_k$ and virtual cavity modes $\hat b_{\mu_k}$. The parameters $\gamma_k=9g$ and $\beta_k=1$. The abbreviations ``$vc$" and ``$co$" of the foot marks stand for initial vacuum and coherent states (similarly in Fig.5 and Fig.9.)}
\label{fig2}
\end{figure}

We consider two kinds of the coupling $g_{\mu_k}(t)$. The first is time-dependent and determined by the most populated modes $\mu_k(t)$ via the relation (\ref{eq3}).  The most populated modes in the output field, which depend on the autocorrelation functions of the cavity modes $\hat b_k$, i.e., 
\begin{align}
\Gamma^{(1)}_k(t_1,t_2)=\gamma_k\langle\hat b_k^\dag(t_1)\hat b_k(t_2)\rangle=\sum\limits_{i}n_{k,i}\mu_{k,i}^*(t_1)
\mu_{k,i}(t_2),
\label{mpm}
\end{align}
where 
$n_{k,i}$ are the mean-photon numbers in each orthogonal  (temporal) modes $\mu_{k,i}(t)$. Here we only consider the most populated mode among the modes $\mu_{k,i}$, which is denoted by $\hat b_{\mu_k}$ in Eq.(\ref{eq2}) with the mode profile $\mu_k$.
The two-time correlation function $\big\langle\hat b_k^\dag(t_1)\hat b_k(t_2)\big\rangle$ can be obtained with the quantum regression theorem and the master equation for the intracavity modes $\hat b_k$ (i.e., $g_{\mu k}=0$ in Eq.(\ref{me})). Fig.\ref{fig2} depicts the modes $\mu_k(t)$ and the time dependence of the mean-photon numbers of $\langle \hat b_k^\dag\hat b_k\rangle$ and $\langle \hat b_{\mu_k}^\dag\hat b_{\mu_k}\rangle$, respectively, for initial vacuum and coherent states of the downconverted cavity modes. Such a consideration gives rise to the time-dependent coupling $g_{\mu_k}$, as shown in Fig.2. Besides, we also consider constant coupling, i.e.,
\begin{align}
g_{\mu_k}=\sqrt{\gamma_{\mu_k}},
\end{align}
which means that
the output field is filtered with generic cavities of Lorentz lineshapes.



\section{Detectable criteria for tripartite non-Gaussian entanglement and steering}
The NTPD process in Eq.(\ref{eq1}) is nonlinear and evolves in non-Gaussian states of which quantum characteristics are determined by various high-order moments. To fully capture non-Gaussian correlation nature in the three-mode system, we introduce single-mode high-order quadratures of the operators $\hat a_k~(a=\{\hat b, \hat b_{\mu}\}~ \text{for the present system})$
\begin{align}
\hat X_k^n =\hat a_k^{\dag n}+\hat a_k^n,~~\hat Y_k^n = i(\hat a_k^{\dag n}-\hat a_k^n)
 \end{align}
for the $k$th mode and two-mode high-order quadratures
\begin{align}
\hat X_{lm}^n =\hat a_l^{\dag n} \hat a_m^{\dag n} + \hat a_l^n \hat a_m^n,~~\hat Y_{lm}^n = i(\hat a_l^{\dag n} \hat a_m^{\dag n} - \hat a_l^n \hat a_m^n),
\end{align}
for the $l$th and $m$th modes $(\{l,m\}=\{1,2,3\})$. The commutation relations for these quadratures $[\hat X_k^n,\hat Y_k^n ]=i\hat C_k^n$ and  $[\hat X_{lm}^n, \hat Y_{lm}^n ]=i\hat C_{lm}^n$. For the present system, $\hat C_k^n=2$ and $8\hat I_k+4$, and
$\hat C_{lm}^n=2(\hat I_l+\hat I_m+1)$ and $4[2+\hat I_l(3+\hat I_l)+\hat I_m^2(1+2\hat I_l)+\hat I_m(3+4\hat I_l+2\hat I_l^2)]$, for $n=1$ and 2, respectively, with $\hat I_k=\hat b_k^\dag \hat b_k$.

To study tripartite entanglement and steering in the system, one can divide the system into a bipartite, i.e., $\{k, (l,m)\}$, and there are three kinds of such bipartition, i.e., $\{1, (2,3)\}$, $\{2, (1,3)\}$ and $\{3, (2,1)\}$. 
The bipartite entanglement between the $k$th mode and the subsystem $(l,m)$ falsifies the separable model for the system's density operator
\begin{align}
\hat\rho_{lm-k}=\sum_{i}\eta_i\hat \rho_{k}^i\hat\rho_{lm}^i,
\label{ec}
\end{align}
where $\sum\limits_{i}\eta_i=1$ and $\hat \rho_{k(l, m)}$ is the density operator of the subsystem $k(l, m)$.
By defining the linear combinations
\begin{align}
\hat U_{k,lm}^n=\hat X_k^n + g_{k,n}\hat X_{lm}^n,~~
\hat V_{k,lm}^n=\hat Y_k^n + h_{k,n}\hat Y_{lm}^n,
\end{align}
with the gain parameters $g_{k,n}$ and $h_{k,n}$, the sum of their variances satisfies
\begin{align}
\big\langle(\Delta\hat U_{k,lm}^n)^2\big\rangle+\big\langle(\Delta\hat V_{k,lm}^n)^2\big\rangle\geq C_k^n+|g_{k,n}h_{k,n}|C_{lm}^n,
\label{ve}
\end{align}
for the bipartite separable model, with $C_{k(lm)}^n=\big\langle\hat C_{k(lm)}^n\big\rangle$. The violation of the above inequality verifies the corresponding bipartite entanglement. The violation of all three inequalities for the three bipartitions, i.e.,
\begin{subequations}
\begin{align}
\big\langle(\Delta\hat U_{1,23}^n)^2\big\rangle+\big\langle(\Delta\hat V_{1,23}^n)^2\big\rangle\geq C_1^n+|g_{1,n}h_{1,n}|C_{23}^n,\\
\big\langle(\Delta\hat U_{2,13}^n)^2\big\rangle+\big\langle(\Delta\hat V_{2,13}^n)^2\big\rangle\geq C_2^n+|g_{2,n}h_{2,n}|C_{13}^n,\\
\big\langle(\Delta\hat U_{3,12}^n)^2\big\rangle+\big\langle(\Delta\hat V_{3,12}^n)^2\big\rangle\geq C_3^n+|g_{3,n}h_{3,n}|C_{12}^n,
\end{align}
\label{var}
\end{subequations}
demonstrates fully inseparable tripartite entanglement. When the whole system is symmetric with respect to the three modes [i.e., the master equation (\ref{me}) is invariant by exchanging the operators $\hat a_1$, $\hat a_2$, $\hat a_3$], the criteria of fully inseparable tripartite entanglement can be simplified as (see the Appendix)
\begin{align}
&\big|\langle\hat a_1^n\hat a_2^n\hat a_3^n\rangle-\langle\hat a_1^n\rangle\langle\hat a_2^n\hat a_3^n\rangle\big|>\nonumber\\
&~~~~~~~~~~~~~\sqrt{\Big[\frac{\big\langle \hat I_1!\big\rangle}{\big\langle \hat I_1^{-n}!\big\rangle}-\langle\hat a_1^n\rangle^2\Big]\Big[\frac{\big\langle \hat I_2!\hat I_3!\big\rangle}{\big\langle \hat I_2^{-n}!\hat I_3^{-n}!\big\rangle}-\langle\hat a_2^n\hat a_3^n\rangle^2\Big]}\nonumber\\
&~~~~~~~~~~~~~\equiv\mathcal {F}_e^n,
\label{fe}
\end{align}
where the sign $\hat I_k^{-n}=(\hat I_k-n)$, with the gain parameters $g_{k,n}=-h_{k,n}$.

The genuine tripartite entanglement is confirmed if the state can not be written as a more general state mixed by the three bipartitions, i.e.,
\begin{align}
\hat\rho_{123}&=P_1\sum\limits_{i_1}\eta_{i_1}\rho_{1}^{i_1}\rho_{23}^{i_1}+P_2\sum\limits_{i_2}\eta_{i_2}\rho_{2}^{i_2}\rho_{13}^{i_2}\nonumber\\
&~~~~+P_3\sum\limits_{i_3}\eta_{i_3}\rho_{3}^{i_3}\rho_{12}^{i_3},
\end{align}
where $\sum_{i}P_i=1$ and $\sum_{i}\eta_i=1$. For the variances in Eq.(\ref{var}), the inequality for confirming the genuine tripartite non-Gaussian entanglement is derived in detail in the Appendix. Again, when the present system is symmetric, the criterion of genuine tripartite non-Gaussian entanglement reduces to (see the Appendix)

\begin{align}
&\big|\langle\hat a_1^n\hat a_2^n\hat a_3^n\rangle-\langle\hat a_1^n\rangle\langle\hat a_2^n\hat a_3^n\rangle\big|>\nonumber\\
&~~~~~~~~~~~~~3\sqrt{\Big[\frac{\big\langle \hat I_1!\big\rangle}{\big\langle \hat I_1^{-n}!\big\rangle}-\langle\hat a_1^n\rangle^2\Big]\Big[\frac{\big\langle \hat I_2!\hat I_3!\big\rangle}{\big\langle \hat I_2^{-n}!\hat I_3^{-n}!\big\rangle}-\langle\hat a_2^n\hat a_3^n\rangle^2\Big]}\nonumber\\
&~~~~~~~~~~~~~\equiv\mathcal {G}_e^n.
\label{ge}
\end{align}

We see that the fully inseparable and genuine tripartite non-Gaussian entanglement depends on the high-order self and cross correlations, i.e.,  $\langle \hat a_k^n\rangle$ and $\big\langle\hat a_l^n\hat a_m^n\big\rangle$. For the present system, when the cavity modes $\hat b_k$ is initially seeded with coherent states, these terms have nonzero values and have obvious effects on the non-Gaussian tripartite entanglement and steering, as will be shown later. When system starts from vacuum or thermal states, $\langle\hat a_k^n\rangle=0$ and $\langle\hat a_l^n\hat a_m^n\rangle=0$, and the above criteria of Eqs.(\ref{fe}) and (\ref{ge}) for $n=1$ are further simplified into
\begin{align}
\big|\langle\hat a_1\hat a_2\hat a_3\rangle\big|>\sqrt{\langle \hat I_1\rangle\langle \hat I_2 \hat I_3\rangle}
\label{fe1}
\end{align}
and
\begin{align}
\big|\langle\hat a_1\hat a_2\hat a_3\rangle\big|>3\sqrt{\langle \hat I_1\rangle\langle \hat I_2 \hat I_3\rangle},
\label{ge1}
\end{align}
respectively, which can also been derived directly with the Hillery-Zubairy entanglement criterion \cite{HZ}.

We next derive the criterion for tripartite steering in the system. Different from the entanglement, the steering of the $k$th mode by the subsystem $(l,m)$ is confirmed by violating the model of local hidden state (LHS), i.e.,
\begin{align}
\hat \rho_{lm\rightarrow k }=\sum\limits_{i}\eta_i\hat \rho_{kQ}^i\rho_{lm}^i,
\end{align}
where we utilize $\hat \rho_{kQ}^i$ and $\hat \rho_{lm}^i$ to replace $\hat \rho_{k}^i$ and $\hat \rho_{lm}^i$ in Eq.(\ref{ec}) respectively, since for the LHS model no explicit assumption is made that $\hat \rho_{lm}^i$ would necessarily be a quantum state described by a quantum density operator. 
According to the LHS model, the sum of the variances of the operators $\hat U_{k,lm}^n$ and $\hat V_{k,lm}^n$ satisfies the inequality
\begin{align}
\big\langle[\Delta\hat U_{k,lm}^n]^2\big\rangle+\big\langle[\Delta\hat V_{k,lm}^n]^2\big\rangle\geq C_k^n,
\label{vs}
\end{align}
whose violation means the bipartite steering from the subsystem $(l, m)$ to the $k$th mode.
The violation of all three inequalities for the three bipartitions
\begin{subequations}
\begin{align}
S_1^n=\big\langle(\Delta\hat U_{1,23}^n)^2\big\rangle+\big\langle(\Delta\hat V_{1,23}^n)^2\big\rangle\geq C_1^n,\\
S_2^n=\big\langle(\Delta\hat U_{2,13}^n)^2\big\rangle+\big\langle(\Delta\hat V_{2,13}^n)^2\big\rangle\geq C_2^n,\\
S_3^n=\big\langle(\Delta\hat U_{3,12}^n)^2\big\rangle+\big\langle(\Delta\hat V_{3,12}^n)^2\big\rangle\geq C_3^n,
\end{align}
\label{str}
\end{subequations}
for any $n$ is sufficient to confirm fully inseparable tripartite steering for the present three-mode system \cite{Run}.
For the symmetric system, the fully inseparable tripartite steering becomes into (see the Appendix)
\begin{align}
&\big|\langle\hat a_1^n\hat a_2^n\hat a_3^n\rangle-\langle\hat a_1^n\rangle\langle\hat a_2^n\hat a_3^n\rangle\big|>\nonumber\\
&~~~~~~~~~~~~~\frac{1}{2}\sqrt{\frac{\big\langle\hat I_2!\hat I_3!\big\rangle}{\big\langle\hat I_2^{-n}!\hat I_3^{-n}!\big\rangle}+\frac{\big\langle\hat I_2^{+n}!\hat I_3^{+n}!\big\rangle}{\big\langle\hat I_2!I_3!\big\rangle}-2\big\langle\hat a_2^n\hat a_3^n\big\rangle^2}\nonumber\\
&~~~~~~~~~~~~~\times\sqrt{\frac{\big\langle\hat I_1!\big\rangle}{\big\langle\hat I_1^{-n}!\big\rangle}
+\frac{\big\langle\hat I_1^{+n}!\big\rangle}{\big\langle\hat I_1!\big\rangle}-\frac{1}{2}C_{1}^n-2\big\langle\hat a_1^n\big\rangle^2}\nonumber\\
&~~~~~~~~~~~~~\equiv\mathcal {F}_s^n.
\label{fs}
\end{align}
with $\hat I_k^{+n}=(\hat I_k+n)$.


Similarly, the genuine tripartite steering is achieved if one can exclude more general
LHS models that are constructed from convex combinations
of LHS models across the three bipartitions \cite{He}, i.e.,
\begin{align}
&\hat \rho_{123}=P_1\sum\limits_{i_1}\eta_{i_1}\hat\rho_{1Q}^{i_1}\rho_{23}^{i_1}+P_2\sum\limits_{i_2}\eta_{i_2}\hat\rho_{2Q}^{i_2}\rho_{13}^{i_2}\nonumber\\
&~~~~~~~~+P_3\sum\limits_{i_3}\eta_{i_3}\hat\rho_{3Q}^{i_3}\rho_{12}^{i_3},
\end{align}
where $\sum\limits_{i}P_i=1$, and $\sum\limits_{i}\eta_i=1$. With Eqs.(\ref{str}), the violation of the inequality
 $S_1^n+S_2^n+S_3^n\geq \text{min}\{C_1^n, C_2^n, C_3^n$\}
for any $n$ is sufficient to certify genuine tripartite non-Gaussian steering.
In our fully symmetric system, the criteria of genuine tripartite non-Gaussian steering can be derived as (see the Appendix)
\begin{align}
&\big|\langle\hat a_1^n\hat a_2^n\hat a_3^n\rangle-\langle\hat a_1^n\rangle\langle\hat a_2^n\hat a_3^n\rangle\big|>\nonumber\\
&~~~~~~~~~~~~~\frac{1}{2}\sqrt{\frac{\big\langle\hat I_2!\hat I_3!\big\rangle}{\big\langle\hat I_2^{-n}!\hat I_3^{-n}!\big\rangle}+\frac{\big\langle\hat I_2^{+n}!\hat I_3^{+n}!\big\rangle}{\big\langle\hat I_2!I_3!\big\rangle}-2\big\langle\hat a_2^n\hat a_3^n\big\rangle^2}\nonumber\\
&~~~~~~~~~~~~~\times\sqrt{\frac{\big\langle\hat I_1!\big\rangle}{\big\langle\hat I_1^{-n}!\big\rangle}
+\frac{\big\langle\hat I_1^{+n}!\big\rangle}{\big\langle\hat I_1!\big\rangle}-\frac{1}{6}C_{1}^n-2\big\langle\hat a_1^n\big\rangle^2}\nonumber\\
&~~~~~~~~~~~~~\equiv\mathcal {G}_s^n.
\label{gs}
\end{align}

For the case of initial thermal or vacuum states, the above criteria of Eqs.(\ref{fs}) and (\ref{gs}) for $n=1$ further reduces to 
\begin{align}
\big|\langle\hat a_1\hat a_2\hat a_3\rangle\big|>\sqrt{\Big\langle (\hat I_2+\frac{1}{2})(\hat I_3+\frac{1}{2})+\frac{1}{4}\Big\rangle\big\langle \hat I_1\big\rangle}
\label{fs1}
\end{align}
and
\begin{align}
\big|\langle\hat a_1\hat a_2\hat a_3\rangle\big|>\sqrt{\Big\langle (\hat I_2+\frac{1}{2})(\hat I_3+\frac{1}{2})+\frac{1}{4}\Big\rangle\big\langle \hat I_1+\frac{1}{3}\big\rangle}.
\label{gs1}
\end{align}
We see that the tripartite non-Gaussian entanglement and steering criteria for $n=1$ just depend on the three-order amplitude correlation $\langle \hat X_1 \hat X_2 \hat X_3\rangle=2\langle\hat a_1\hat a_2\hat a_3\rangle$ for the present system, intensity correlations $\langle\hat I_k\hat I_{k'}\rangle$ and intensities $\langle\hat I_k\rangle$, which can be measured in the recent NTPD experiment \cite{Sandbo}. Note that it is shown from Eqs.(\ref{fe1}) and (\ref{fs1}) that the condition for achieving the full inseparable tripartite steering is more strict than that for the full inseparable tripartite entanglement, but it is not true for the genuine tripartite entanglement and steering, as revealed by Eqs.(\ref{ge1}) and (\ref{gs1}). This is essentially because that these conditions are sufficient for detecting the entanglement and steering. Note that in deriving the inequality (\ref{a6}), six terms on the right hand in the inequality (\ref{a4}) are discarded, different from the derivation of the condition for genuine tripartite entanglement in Eq.(\ref{a11}).

\section{Numerical results}
In this section, we investigate in detail the features of intracavity and output non-Gaussian tripartite entanglement and steering in the NTPD. We define the quantities $E_{f(g)}^n=\big|\langle\hat a_1^n\hat a_2^n\hat a_3^n\rangle-\langle\hat a_1^n\rangle\langle\hat a_2^n\hat a_3^n\rangle\big|-\mathcal {F(G)}_e^n$ and $S_{f(g)}^n=\big|\langle\hat a_1^n\hat a_2^n\hat a_3^n\rangle-\langle\hat a_1^n\rangle\langle\hat a_2^n\hat a_3^n\rangle\big|-\mathcal {F(G)}_s^n$ to characterize the full inseparable (genuine) tripartite entanglement and steering. Their existences are signified by the conditions $S_{f(g)}^n>0$ and $E_{f(g)}^n>0$.
\begin{figure}[t]
\vspace{0cm}
\centerline{\hspace{-0.2cm}\scalebox{0.24}{\includegraphics{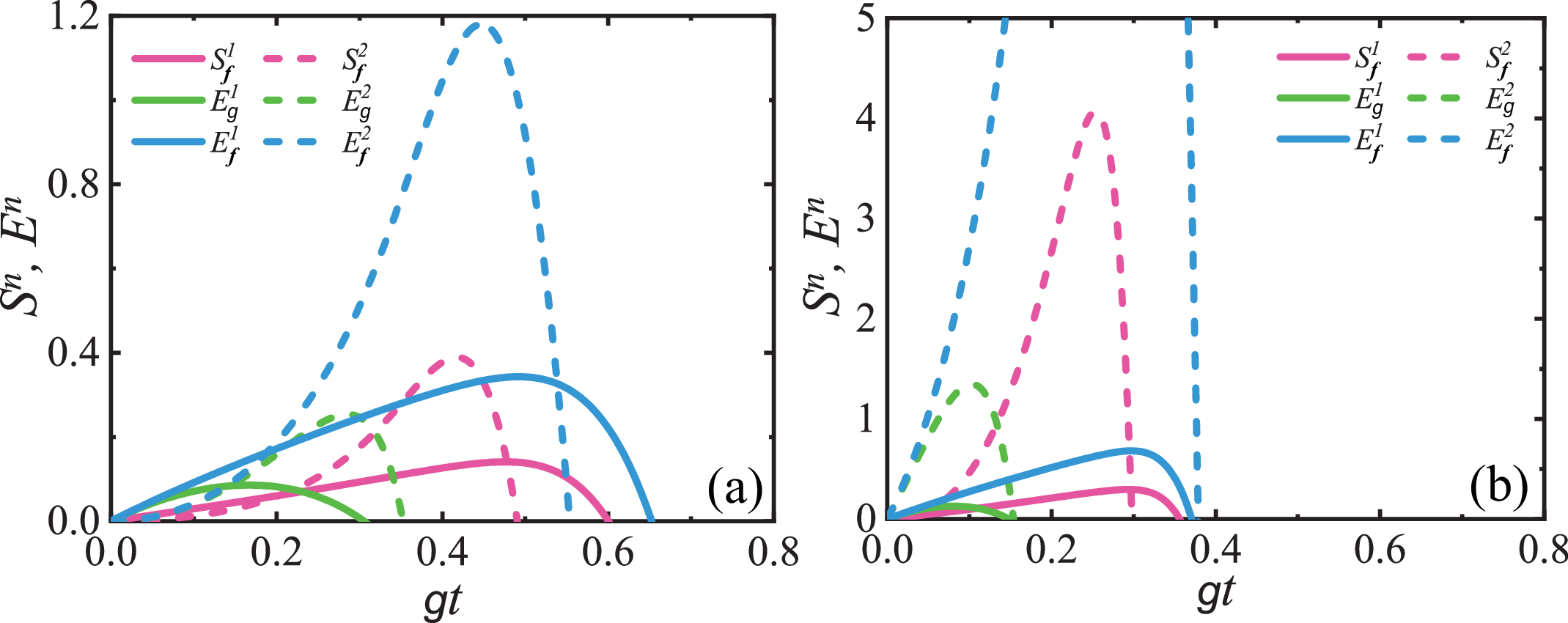}}}
\vspace{0cm}
\caption{The time evolution of the tripartite entanglement $E^n$ and steering $S^n(n=1,2)$ inside the cavity for $\gamma_k=0$. The initial states of the cavity modes are considered to be vacua in (a) and coherent states in (b) with the amplitude $\beta_k=1$.}
\label{fig3}
\end{figure}
\begin{figure}[t]
\vspace{0cm}
\centerline{\hspace{-0.2cm}\scalebox{0.24}{\includegraphics{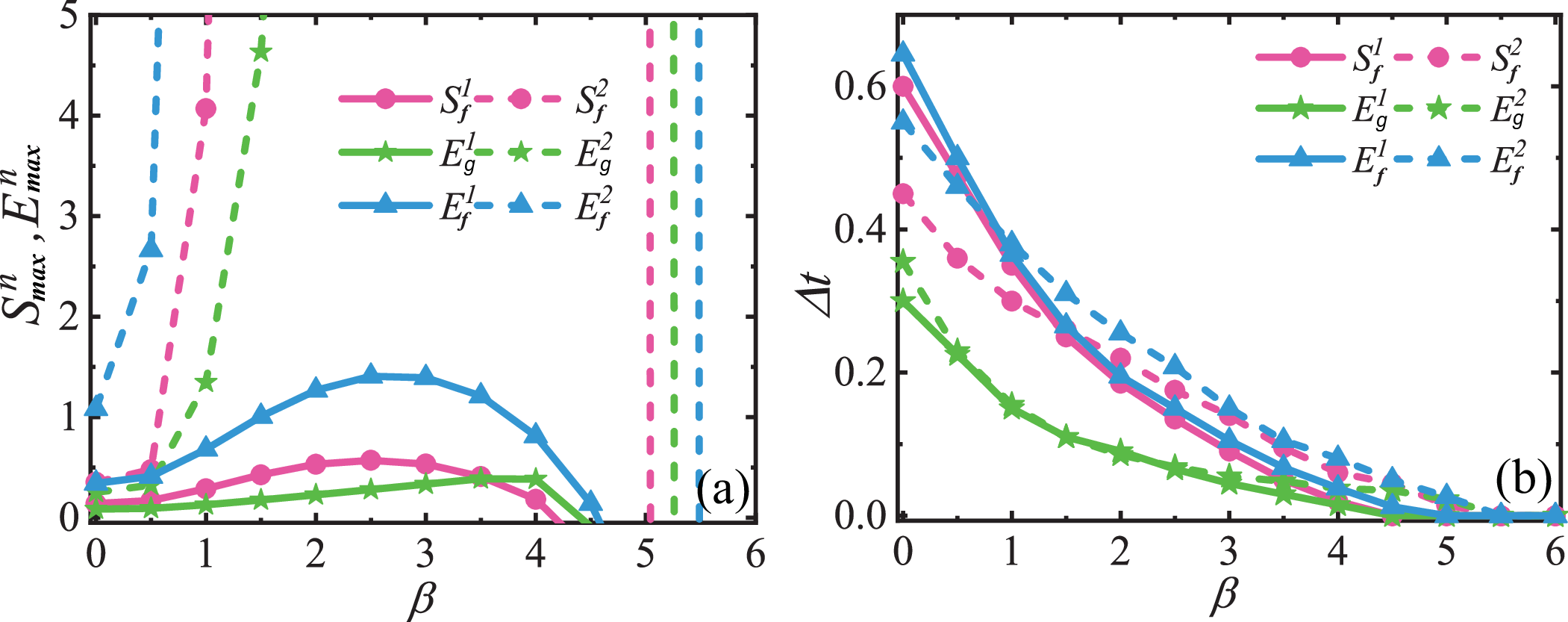}}}
\vspace{0cm}
\caption{(a) The dependence of the maximum entanglement $E_{max}^n$ and steering $S_{max}^n$ inside the cavity on the amplitude $\beta_k$ of the initial cavity-field coherent states. (b) The duration $\Delta t$ of the entanglement and steering versus the coherent amplitude $\beta_k$. The parameter are the same as Fig.\ref{fig3}.}
\label{fig4}
\end{figure}
\begin{figure}[t]
\vspace{0cm}
\centerline{\hspace{-0.3cm}\scalebox{0.255}{\includegraphics{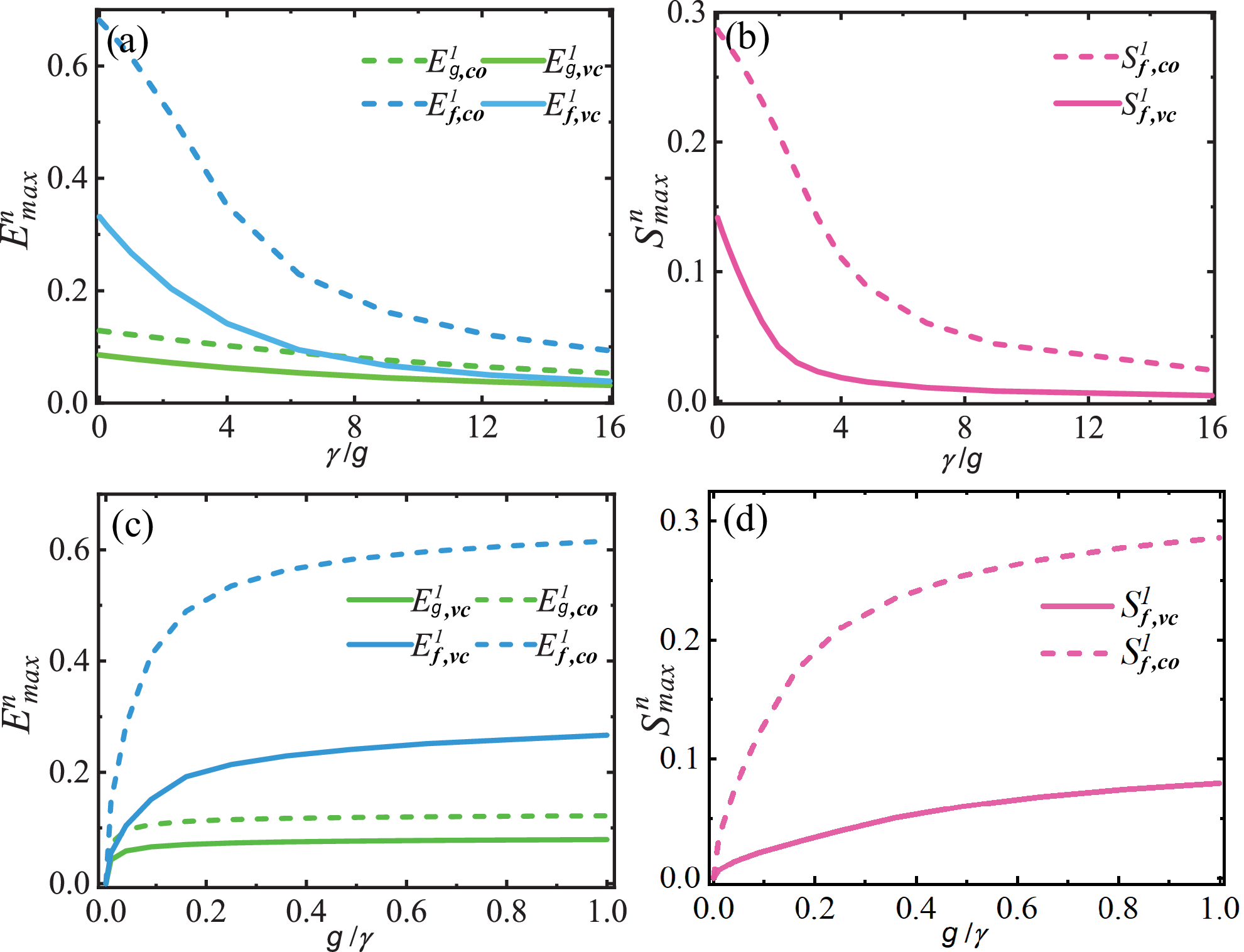}}}
\vspace{0cm}
\caption{The maximum entanglement $E_{max}^n$ and steering $S_{max}^n~(n=1)$ inside the cavity as the functions of the cavity dissipation rates $\gamma_k=\gamma$ [(a) and (b)] and the coupling $g$ [(c) and (d)], for the initial cavity-field coherent amplitude $\beta_k=1$. }
\label{fig5}
\end{figure}

In Fig.3, the time evolution of the tripartite entanglement and steering inside the cavity are plotted for $n=1$ and $n=2$, with initial vacuum and coherent states of the cavity modes $\hat b_k$. It is shown that the non-Gaussian tripartite genuine entanglement and fully inseparable steering can be achieved in the short-time regime. The genuine tripartite steering cannot be found (at least) with the quantity $S_{g}^n$. It is obviously shown that for the same value of $n$, the full inseparable entanglement lasts longer than genuine entanglement, since the latter exhibits stronger quantum correlations, and both of them for $n=1$ last longer than those for $n=2$, which is contrary to the case of the steering. Compared to the case of initial vacuum states, the maximally-achievable tripartite entanglement and steering are obviously enhanced by initial coherent states, as shown in Fig.3~(b). The reason may be that for the NTPD process, the initial coherent seeding of the cavity modes results in nonzero single-mode and two-mode high-order correlations, i.e., $\langle \hat b_k^n\rangle\neq0$ and $\langle \hat b_l^n\hat b_m^n\rangle\neq0$, which in turn increase the tripartite non-Gaussian quantum correlations. Fig.4 (a) plots the dependence of the maximal tripartite entanglement and steering on the amplitudes $\beta_k=\beta$, and it shows that the increasing of the initial amplitudes $\beta$, the maximal tripartite entanglement and steering  increases first, then decrease, and finally disappear. This is because that in this situation we can express the downconverted cavity mode $\hat b_k$ as the sum of average amplitude $\langle \hat b_k\rangle$ and corresponding quantum fluctuation $\delta \hat b_k$ around the amplitude. Then, the Hamiltonian $\hat H_s$ in Eq.(\ref{eq1}) can be divided into two parts: the linearized and nonlinear parts which are respectively dependent and independent on the average amplitude $\langle \hat b_k\rangle$. As the initial amplitude $\beta$ increases such that the linearized part dominates over the nonlinear one, the NTPD mainly appears as a Gaussian system, i.e., three concurrent two-mode nondegenerate parametric downconversion of $\delta \hat b_k$ and its non-Gaussian characteristics is suppressed, which leads to that the non-Gaussian entanglement and steering decrease gradually and the NTPD dominantly exhibits Gaussian tripartite entanglement and steering. We therefore see that the initial preparation of the downconverted modes in weak coherent states is helpful to the generation of the tripartite non-Gaussian entanglement and steering. In addition, it shows from Fig.4 (b) that the initial coherent states shorten the existence time of the tripartite entanglement and steering. Fig.5 illustrates the dependence of the maximal tripartite entanglement and steering on the cavity dissipation rates $\gamma_k$ and interaction strength $g$. As expected, the tripartite entanglement and steering decrease with the increasing of the dissipation rates, and they eventually disappear when the dissipation rates $\gamma_k\gg g$, irrespective of initial vacuum or coherent states, as plotted in Fig.5~(a) and (b), since the intracavity field escapes rapidly from the cavity for the large cavity dissipation rates. 
In addition, it can be seen from Fig.5 (c) and (d) that the maximal tripartite entanglement and steering increase as the interaction strength $g$ increases and the growth rates decrease gradually when the strength further arises.
\begin{figure}[t]
\centerline{\hspace{-0.3cm}\scalebox{0.295}{\includegraphics{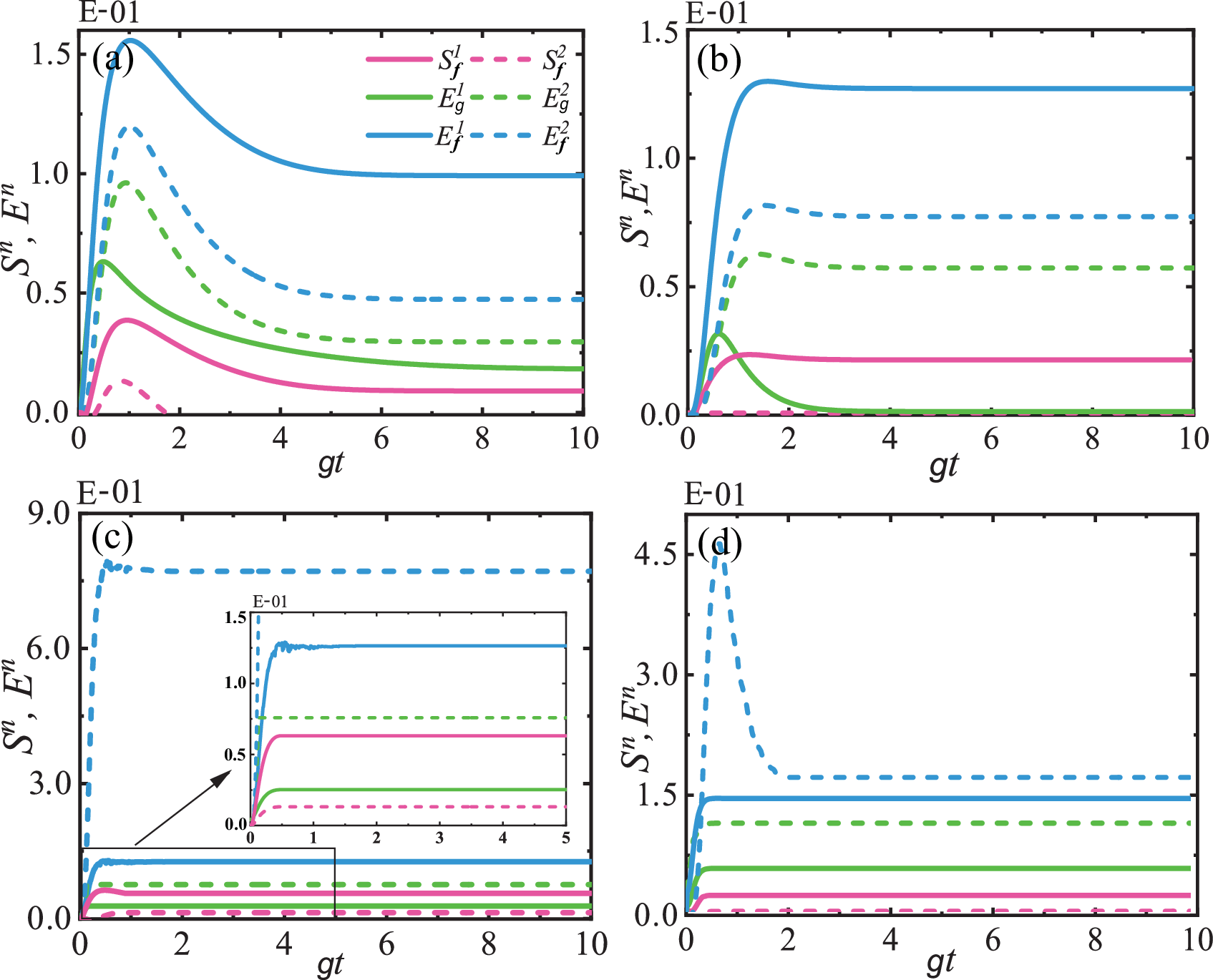}}}
\caption{The time evolution of the entanglement $E^n$ and steering $S^n~(n=1,2)$ of the cavity output field for initial cavity-field vacua [(a) and (b)] and coherent states [(c) and (d)] with the amplitude $\beta_k=1$, with the time-dependent coupling $g_{\mu_k}(t)$ [(a) and (c)] and constant coupling $g_{\mu_k}(t)=1.5\sqrt{g}$ [(b)] and (d)]. The cavity dissipation rates $\gamma_k=9g$.}
\label{fig6}
\end{figure}



We next investigate the properties of the tripartite non-Gaussian entanglement and steering in the output field by investigating the virtual cavity modes $\hat b_{\mu_k}$.  In Fig.6, the time evolution of the entanglement and steering is plotted for the time-dependent and constant couplings $g_{\mu_k}$, with initial vacuum and coherent states of the intracavity modes. It shows that steady-state non-Gaussian genuine tripartite entanglement and fully inseparable tripartite steering can be achieved, although they are just present in the short-time regime inside the cavity. This can be understood as that the variances in Eqs.(\ref{ve}) and (\ref{vs}) of the intracavity field can be considered as the sum of those of all output modes $\mu_k$ and therefore the steady-state tripartite entanglement and steering in output field may be generated, although they only exist in a finite time. 
Further, we see from Fig.6 ~(a) and (c) that the steady-state tripartite entanglement and steering for $n=1$ and $2$ can also be enhanced by the initial coherent states of the downconverted cavity modes. The entanglement and steering for the initial vacua in Fig.6 (a) drop and slowly stabilize after reaching the maximal values, while they stabilize as the maxima are reached for the case of initial coherent states in Fig.6 (c), as the coupling $g_{\mu_k}$ [see Fig.2 (a)] approaches the steady states much faster in the latter case.
The entanglement and steering for the constant coupling $g_{\mu_k}$ are less improved with the coherent states in Fig.6 (d), but much faster reach the steady states compared to the case of initial vacua in Fig.6 (b).
In Fig.6 (c) and (d), the entanglement $E^1_g$ and $E^2_f$ go down after reaching the highest points because that the ratio of the cavity dissipation rates $\sqrt{\gamma_k}$ to the constant coupling $g_{\mu_k}$ is not optimized for them and here the same ratio is settled simply.
The purity of the output states, defined by $P=\text{Tr}[(\hat \rho_{b_{\mu 1}b_{\mu 2}b_{\mu 3}})^2]$, is plotted in Fig.7. It is shown that the purity for the initial coherent states is decreased, although the entanglement and steering are enhanced by them, compared to the case of the initial vacuum states.
In addition, the purity for the constant coupling in Fig.6 (b) and (d) is obviously higher than that in Fig.6 (a) and (c) because of the larger coupling $g_{\mu_k}$.


\begin{figure}[t]
\vspace{0cm}
\centerline{\hspace{-0.5cm}\scalebox{0.22}{\includegraphics{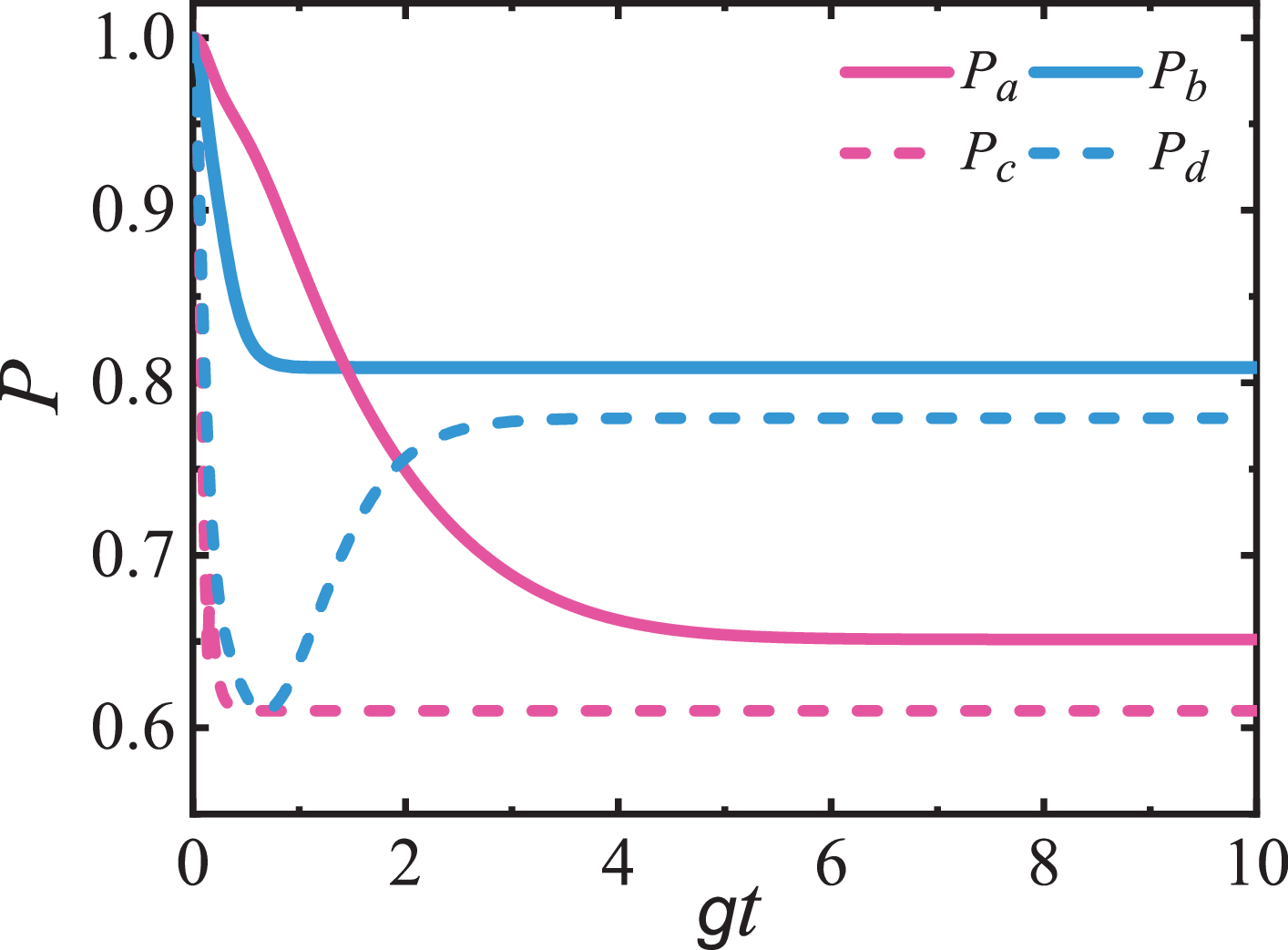}}}
\vspace{0.0cm}
\caption{The purity $P_{a}-P_{d}$ of the cavity output states $\hat \rho_{b_{\mu 1}b_{\mu 2}b_{\mu 3}}$ corresponding to the states in Fig.6 (a)-(d), respectively.}
\label{fig7}
\end{figure}




We finally consider the application of the steady-state output tripartite non-Gaussian steering to remotely generating negative Wigner-function conditional states via homodyne detection. Specifically, we investigate the conditional states of the output mode $\hat b_{\mu_3}$ by homodying the quadratures $\hat X_{b_{\mu_{1(2)}}}=(\hat b_{\mu_{1(2)}}+\hat b_{\mu_{1(2)}}^\dag)$ of the output modes $\hat b_{\mu_{1}}$ and $\hat b_{\mu_{2}}$. Conditioned on the homodyne detection outcomes $x_{b_{\mu_{1}}}$ and $x_{b_{\mu_{2}}}$, the density operator of the mode $\hat b_{\mu_{3}}$ \cite{MW}
\begin{align}
\hat \rho_{b_{\mu_{3}}}(x_{b_{\mu_{1}}},x_{b_{\mu_{2}}})=\frac{\hat {\tilde{\rho}}_{b_{\mu_{3}}}(x_{b_{\mu_{1}}},x_{b_{\mu_{2}}})}{\text{Tr}_{b_{\mu_{3}}} \big[\hat {\tilde{\rho}}_b(x_{b_{\mu_{1}}},x_{b_{\mu_{2}}})\big]},
\label{e3}
\end{align}
where the unnormalzied operator $\hat {\tilde{\rho}}_{b_{\mu_{3}}}(x_{b_{\mu_{1}}},x_{b_{\mu_{2}}})=\text{Tr}_{b_{\mu_{1}}b_{\mu_{2}}} \big[(\hat {\mathcal M}_{b_{\mu_{1}}b_{\mu_{2}}} \otimes \hat I_{b_{\mu_{3}}})\hat \rho_{b_{\mu_{1}}b_{\mu_{2}}b_{\mu_{3}}}(\hat I_{b_{\mu_{3}}}\otimes\hat {\mathcal M}_{b_{\mu_{1}}b_{\mu_{2}}})\big]$ and the projection operator $\hat {\mathcal M}_{b_{\mu_{1}}b_{\mu_{2}}}=|{b_{\mu_{1}},b_{\mu_{2}}}\rangle\langle {b_{\mu_{1}},b_{\mu_{2}}}|$, which can be calculated in the Fock space with $\langle x_{o}\mid n_{}\rangle=\frac{1}{\pi^{1/4}}\frac{1}{\sqrt{2^{n_{o}}n_{o}!}}e^{-x_{{o}}^{2}/2}H_{n_{o}}(x_{o})$, with $H_{n_{o}}$ being the Hermite polynomial of order $n_{o}$ and $o=\{b_{\mu_{1}}, b_{\mu_{2}}\}$.

\begin{figure}[t]
\vspace{0cm}
\centerline{\hspace{-0.2cm}\scalebox{0.32}{\includegraphics{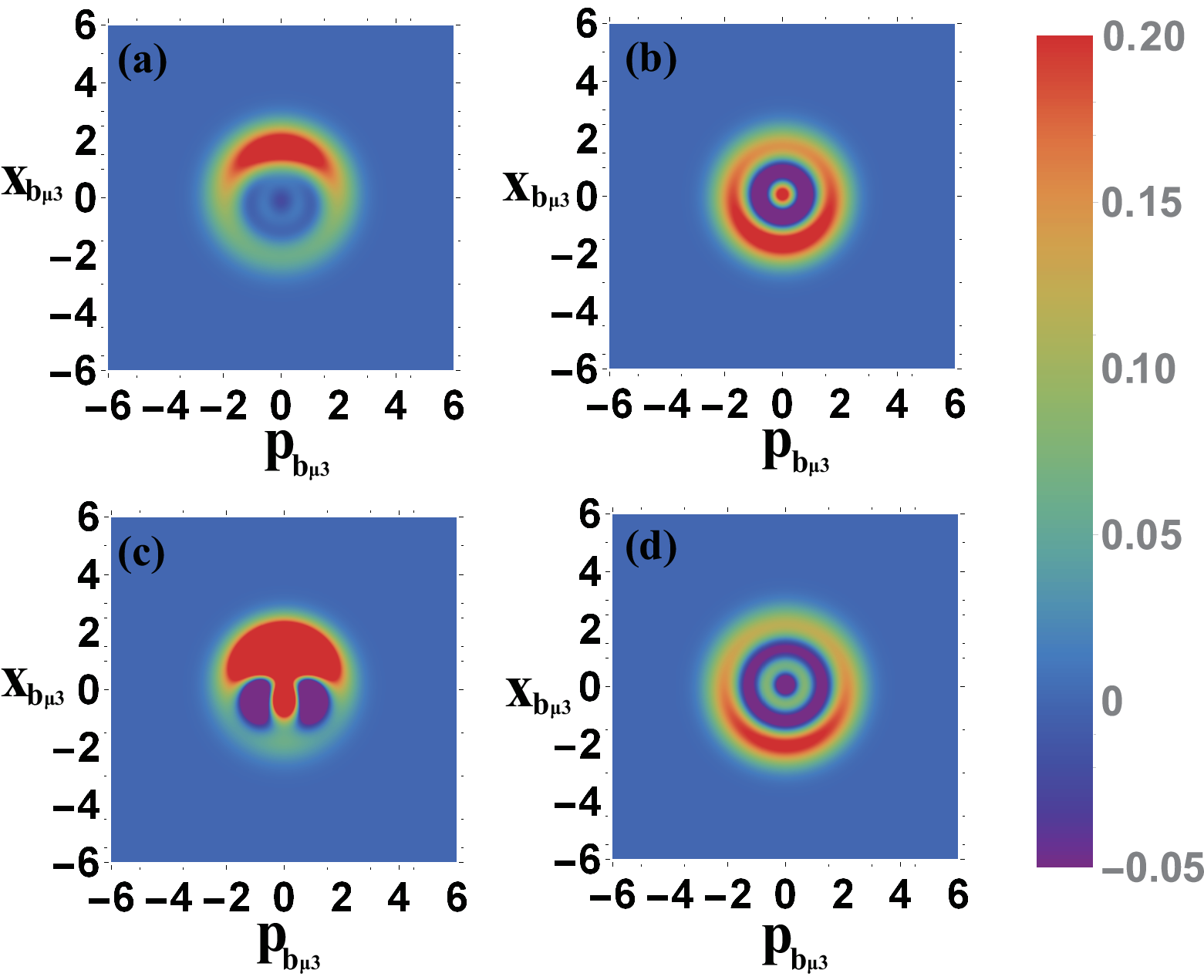}}}
\vspace{0cm}
\caption{The density plots of the Wigner functions $W_{b_{\mu_{3}}}(x_{b_{\mu_{3}}},p_{b_{\mu_{3}}})$ of the conditioned final state $\hat \rho_{b_{\mu_{3}}}(x_{{b_{\mu_{1}}}}=x_{{b_{\mu_{2}}}}=5)$ corresponding to the states in Fig.6 (a)-(d), respectively.}
\label{fig8}
\end{figure}

In Fig.8 (a)-(d), the density plots of the Wigner functions $W_{b_{\mu_{3}}}(x_{b_{\mu_{3}}},p_{b_{\mu_{3}}})$, obtained by performing Fourier transform on the characteristic function defined via $\chi_{b_{\mu_{3}}}(\xi)=\text{Tr}\big[e^{\xi\hat b_{\mu_{3}}^\dag -\xi^*\hat b_{\mu_{3}}}\hat \rho_{b_{\mu_{3}}}(x_{b_{\mu_{1}}},x_{b_{\mu_{2}}})\big]$, are presented for the tripartite non-Gaussian steerable states Fig.6 (a)-(d), respectively, with the homodyne detection outcomes $x_{{b_{\mu_{1}}}}=x_{{b_{\mu_{2}}}}=5$. It shows that the Wigner functions exhibits negativity
\begin{align}
\mathcal N=\int\Big[\big|W(\alpha,\alpha^{*})\big|-W(\alpha,\alpha^{*})\Big]d^{2}\alpha,
\end{align}
with phase-space variable $\alpha=x_{b_{\mu_{3}}}+ip_{b_{\mu_{3}}}$, indicating genuine non-Gaussian nonclassicality.
 Essentially, the capability for this remote generation of negative Wigner states is endowed with the non-Gaussian steerable nonlocality generated in the NTPD process.
\begin{figure}[t]
\vspace{0cm}
\centerline{\hspace{-0.25cm}\scalebox{0.233}{\includegraphics{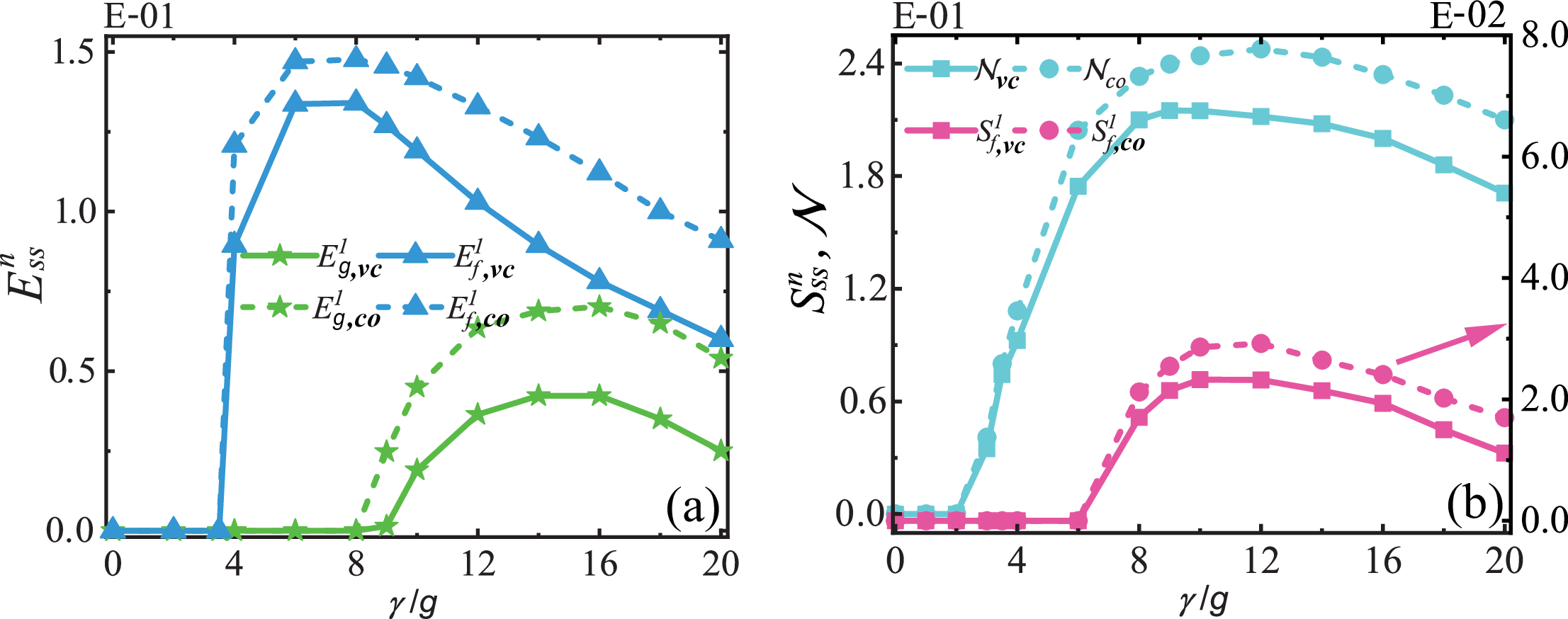}}}
\vspace{0cm}
\caption{(a) The steady-state tripartite entanglement $E^n_{ss}$ for the constant coupling $g_{\mu_k}=1.5\sqrt{g}$ (initial vacua and coherent states with the amplitude $\beta_k=1$) as the function of the cavity dissipation rate $\gamma_k$. (b) The same for the steady-state tripartite steering $S^n_{ss}$ and the Wigner negativity $\mathcal N$ of the conditioned final state $\hat \rho_{b_{\mu_{3}}}(x_{{b_{\mu_{1}}}}=x_{{b_{\mu_{2}}}}=5)$.}
\label{fig9}
\end{figure}
In Fig.9, we shown the effects of the cavity dissipation rates $\gamma_k$ on the steady-state tripartite entanglement, steering and Wigner negativity for the case of the constant coupling $g_{\mu_k}$ in Fig.6 (b) and (d). In Fig.9 (a), the fully inseparable tripartite entanglement increases rapidly first and then decreases with the dissipation rates, but the genuine tripartite entanglement just increases only after the dissipation rate reaches a certain value, due to the fact that the latter exhibits  stronger correlations than the former. As the dissipation rate increases, the dissipative cascaded coupling increases and the entanglement thus increases accordingly, and as the dissipation rate continues to increase, the entanglement and steering are decreased by the dissipation. In Fig.9 (b), the Wigner negativity and the fully insepaprable tripartite steering increases rapidly and then decreases slowly over the range of the dissipation rate. It is shown clearly that the negativity has the similar dependence on the dissipation rate to that of the tripartite steering and  the improved steering for initial coherent states gives larger negativity, which therefore reflects the intrinsic capability of quantum steering for manipulating local quantum states via remote detection.

\section{Conclusion}
In summary, we study in this paper the properties of tripartite non-Gaussian entanglement and steering in an intracavity NTPD process. We derive the criteria for fully inseparability and genuine tripartite non-Gaussian entanglement and steering with high-order field quadratures for the present system. With the criteria and visualizing the specific modes in the output continuous field as virtual cavities coupled to the NTPD cavity in a cascade way, the tripartitie non-Gaussian entanglement and steering inside and outside the cavity are studied in detailed. It is found that the tripartite non-Gaussian entanglement and steering inside the cavity only exist in the short-time regime but they can be generated in the steady-state regime in the output field of the cavity. Moreover, it is revealed that the initial coherent cavity-field states can effectively enhance the output steady-state and intracavity transient entanglement and steering.
It is also shown that the output tripartite non-Gaussian steering can be utilized to remotely generate non-Gaussian states with negative Wigner functions by homodyne detection. Our findings unravel the novel non-Gaussian nonclassical characteristics in the nonlinear NTPD process. Further work may include the investigation on the application of the output triple photons in genuine tripartite non-Gaussian entangled states to various quantum tasks, such as quantum parameter estimation and quantum illumination.

\section*{ACKNOWLEDGMENT}
This work is supported by the National Natural Science Foundation of China (Grant No.12174140).


\setcounter{equation}{0}
\renewcommand\theequation{A\arabic{equation}} 
\section*{APPENDIX: DERIVATION OF THE NON-GAUSSIAN TRIPARTITE ENTANGLEMENT AND STEERING}
In the paper, the high-order quadrature operators are defined as $\hat X_k^n =(\hat a_k^{\dag n}+\hat a_k^n)$ and $\hat Y_k^n = i(\hat a_k^{\dag n}-\hat a_k^n)$, where
$\hat a_k$ and $\hat a_k^{\dag}$
are the annihilation and creation operators with $[\hat a_k,\hat a_k^{\dag}]=1$. $\hat X_k^n$ and $\hat Y_k^n$ satisfy the commutation relation $[\hat X_k^n,\hat Y_k^n]=i\hat C_k^n$. $\hat X_{lm}^n$ and $\hat Y_{lm}^n$ satisfy the commutation relation $[\hat X_{lm}^n,\hat Y_{lm}^n]=i\hat C_{lm}^n$.

According to the biseparable state $\rho_{k,lm}=\sum\limits_{i}\eta_i\rho_{k}^i\rho_{lm}^i$, the total variance of the pair of operators $\hat U_{k,lm}^n$ and $\hat V_{k,lm}^n$ satisfies the inequality
\begin{widetext}
\begin{equation}
\begin{split}
&~~~~\langle(\Delta\hat U_{k,lm}^n)^2\rangle+\langle(\Delta\hat V_{k,lm}^n)^2\rangle\\&=\langle(\hat U_{k,lm}^n)^2\rangle+\langle(\hat V_{k,lm}^n)^2\rangle-\langle\hat U_{k,lm}^n\rangle^2-\langle\hat V_{k,lm}^n\rangle^2\\&
=\sum\limits_{i}\eta_i\{\langle(\hat X_k^n)^2+(g_{k,n}\hat X_{lm}^n)^2+(\hat Y_k^n)^2+(h_{k,n}\hat Y_{lm}^n)^2\rangle_i+2g_{k,n}\langle\hat X_k^n\hat X_{lm}^n\rangle_i+2h_{k,n}\langle\hat Y_k^n\hat Y_{lm}^n\rangle_i\}-\langle\hat U_{k,lm}^n\rangle^2-\langle\hat V_{k,lm}^n\rangle^2\\&
=\sum\limits_{i}\eta_i\{\langle(\Delta\hat X_k^n)^2+(\Delta\hat Y_k^n)^2+(g_{k,n}\Delta\hat X_{lm}^n)^2+(h_{k,n}\Delta\hat Y_{lm}^n)^2\rangle_i+2g_{k,n}(\langle\hat X_k^n\hat X_{lm}^n\rangle_i-\langle\hat X_k^n\rangle_i\langle\hat X_{lm}^n\rangle_i)\\&+2h_{k,n}(\langle\hat Y_k^n\hat Y_{lm}^n\rangle_i-\langle\hat Y_k^n\rangle_i\langle\hat Y_{lm}^n\rangle_i)\}+\sum\limits_{i}\eta_i\langle\hat U_k^n\rangle_i^2-(\sum\limits_{i}\eta_i\langle\hat U_k^n\rangle_i)^2+\sum\limits_{i}\eta_i\langle\hat V_k^n\rangle_i^2-(\sum\limits_{i}\eta_i\langle\hat V_k^n\rangle_i)^2\\&\geq\sum\limits_{i}\eta_i\{\langle(\Delta\hat X_k^n)^2+(\Delta\hat Y_k^n)^2+(g_{k,n}\Delta\hat X_{lm}^n)^2+(h_{k,n}\Delta\hat Y_{lm}^n)^2\rangle_i\}
\\&\geq C_k^n+|g_{k,n}h_{k,n}|C_{lm}^n
\end{split}
\end{equation}
\end{widetext}
In the derivation process, we utilized  the Cauchy-Schwartz inequality $\sum\limits_{i}\eta_i\langle\hat U^n\rangle_i^2 \geq (\sum\limits_{i}\eta_i\langle\hat U^n\rangle_i)^2$
the sum uncertainty relation $\langle(\Delta\hat X^n)^2\rangle+\langle(\Delta\hat Y^n)^2\rangle\geq |\langle[\hat X^n,\hat Y^n]\rangle|$ and $\langle\hat C^n\rangle\equiv C^n$.

But if the state of subsystems $l$ and $m$ is not assumed to be a quantum state, there is only the assumption of non-negativity for the associated variances. For the biseparable local hidden state model symbolized as $\rho_{lm\rightarrow k}=\sum\limits_{i}\eta_i\rho_{kQ}^i\rho_{lm}^i$,
\begin{widetext}
\begin{equation}
\begin{split}
\langle(\Delta\hat U_{k,lm}^n)^2\rangle+\langle(\Delta\hat V_{k,lm}^n)^2\rangle&=\langle[\Delta(\hat X_k^n + g_{k,n}\hat X_{lm}^n)]^2\rangle+\langle[\Delta(\hat Y_k^n + h_{k,n}\hat Y_{lm}^n)]^2\rangle\\&
\geq\sum\limits_{i}\eta_i\{\langle(\Delta\hat X_k^n)^2\rangle_i+\langle\Delta(g_{k,n}\hat X_{lm}^n)^2\rangle_i+\langle(\Delta\hat Y_k^n)^2\rangle_i+\langle\Delta(h_{k,n}\hat Y_{lm}^n)^2\rangle_i\}\\&
\geq\sum\limits_{i}\eta_i\{\langle(\Delta\hat X_k^n)^2\rangle_i+\langle(\Delta\hat Y_k^n)^2\rangle_i\}
\geq C_k^n
\end{split}
\end{equation}
\end{widetext}

In the derivation process, we utilized  the nonnegativity of variances that for any local hidden variables, i.e, $\langle\Delta(g_{k,n}\hat X_{lm}^n)^2\rangle\geq 0$ and $\langle\Delta(h_{k,n}\hat Y_{lm}^n)^2\rangle\geq 0$,
the sum uncertainty relation $\langle(\Delta\hat X_{k}^n)^2\rangle+\langle(\Delta\hat Y_{k}^n)^2\rangle\geq |\langle[\hat X_k^n,\hat Y_k^n]\rangle|$ and $\langle\hat C_k^n\rangle\equiv C_k^n$.

Then, the three inequalities can be written as
\begin{equation}
\begin{split}
S_1^n=\langle(\Delta\hat U_{1,23}^n)^2\rangle+\langle(\Delta\hat V_{1,23}^n)^2\rangle\geq C_1^n,\\
S_2^n=\langle(\Delta\hat U_{2,13}^n)^2\rangle+\langle(\Delta\hat V_{2,13}^n)^2\rangle\geq C_2^n,\\
S_3^n=\langle(\Delta\hat U_{3,12}^n)^2\rangle+\langle(\Delta\hat V_{3,12}^n)^2\rangle\geq C_3^n,
\end{split}
\end{equation}
Violation of the three inequalities above can confirm fully inseparable tripartite steering.

In our symmetric system, $S_1^n=S_2^n=S_3^n, C_1^n=C_2^n=C_3^n, g_{1,n}=g_{2,n}=g_{3,n}, h_{1,n}=h_{2,n}=h_{3,n}$, i.e., violating any of the above formulas can confirm fully inseparable tripartite steering. Calculating
\begin{equation}
\begin{split}
S_1^n&=\langle(\Delta\hat U_{1,23}^n)^2\rangle+\langle(\Delta\hat V_{1,23}^n)^2\rangle\\&=\langle(\hat X_{1}^n)^2\rangle-\langle\hat X_{1}^n\rangle^2+2g_{1,n}(\langle\hat X_{1}^n\hat X_{23}^n\rangle-\langle\hat X_{1}^n\rangle\langle\hat X_{23}^n\rangle)\\&+g_{1,n}^2(\langle(\hat X_{23}^n)^2\rangle-\langle\hat X_{23}^n\rangle^2)+\langle(\hat Y_{1}^n)^2\rangle-\langle\hat Y_{1}^n\rangle^2\\&+2h_{1,n}(\langle\hat Y_{1}^n\hat Y_{23}^n\rangle-\langle\hat Y_{1}^n\rangle\langle\hat Y_{23}^n\rangle)\\&+h_{1,n}^2(\langle(\hat Y_{23}^n)^2\rangle-\langle\hat Y_{23}^n\rangle^2),
\end{split}
\end{equation}
where the optimal gain parameters
$g_{1,n}=-h_{1,n}=\frac{-(\langle\hat X_{1}^n\hat X_{23}^n\rangle-\langle\hat X_{1}^n\rangle\langle\hat X_{23}^n\rangle)+(\langle\hat Y_{1}^n\hat Y_{23}^n\rangle-\langle\hat Y_{1}^n\rangle\langle\hat Y_{23}^n\rangle)}{(\langle(\hat X_{23}^n)^2\rangle-\langle\hat X_{23}^n\rangle^2)+(\langle(\hat Y_{23}^n)^2\rangle-\langle\hat Y_{23}^n\rangle^2)}$. Bringing the parameters back to the first formula in Eq.(A3), we can get a simplified inequality
\begin{align}
&\big|\langle\hat a_1^n\hat a_2^n\hat a_3^n\rangle-\langle\hat a_1^n\rangle\langle\hat a_2^n\hat a_3^n\rangle\big|\leq\nonumber\\
&~~~~~~\frac{1}{2}\sqrt{\big\langle\hat a_2^{\dag n}\hat a_2^n\hat a_3^{\dag n}\hat a_3^n\big\rangle+\big\langle\hat a_2^n\hat a_2^{\dag n}\hat a_3^n\hat a_3^{\dag n}\big\rangle-2\big\langle\hat a_2^n\hat a_3^n\big\rangle^2}\nonumber\\
&~~~~~~\times\sqrt{\big\langle\hat a_1^{\dag n}\hat a_1^n\big\rangle+\big\langle\hat a_1^n\hat a_1^{\dag n}\big\rangle-\frac{1}{2}C_{1}^n-2\big\langle\hat a_1^n\big\rangle^2}.
\end{align}
Violation of the inequality above can confirm fully inseparable tripartite steering.

Futhermore, we consider that the system is described by mixtures of the type
$\rho_{mix}=P_1\sum\limits_{i_1}\eta_{i_1}\rho_{1Q}^{i_1}\rho_{23}^{i_1}+P_2\sum\limits_{i_2}\eta_{i_2}\rho_{2Q}^{i_2}\rho_{13}^{i_2}+P_3\sum\limits_{i_3}\eta_{i_3}\rho_{3Q}^{i_3}\rho_{12}^{i_3}$,
where $\sum\limits_{i}P_i=1$ and $\sum\limits_{i}\eta_i=1$. Then substituting the mixture state into Eq.(A3), we find
\begin{equation}
\begin{split}
S_1^n\geq P_{1}S_{1,1}^n+P_{2}S_{1,2}^n+P_{3}S_{1,3}^n\\
S_2^n\geq P_{1}S_{2,1}^n+P_{2}S_{2,2}^n+P_{3}S_{2,3}^n,\\
S_3^n\geq P_{1}S_{3,1}^n+P_{2}S_{3,2}^n+P_{3}S_{3,3}^n,
\label{a4}
\end{split}
\end{equation}
where $S_{j,j'}^n$ stands for the total
variance of operators $\hat U_j$ and $\hat V_j$ over the density operator $\rho_j$. Thus we have
\begin{widetext}
\begin{equation}
\begin{split}
S_{1,1}^n&=\langle(\Delta\hat U_{1,23}^n)^2\rangle_1+\langle(\Delta\hat V_{1,23}^n)^2\rangle_1\\&=\langle[\Delta(\hat X_1^n + g_{1,n}\hat X_{23}^n)]^2\rangle_1+\langle[\Delta(\hat Y_1^n + h_{1,n}\hat Y_{23}^n)]^2\rangle_1\\&
\geq P_1\sum\limits_{i_1}\eta_{i_1}\{\langle(\Delta\hat X_1^n)^2\rangle_{i_1}+\langle\Delta(g_{1,n}\hat X_{23}^n)^2\rangle_{i_1}+\langle(\Delta\hat Y_1^n)^2\rangle_{i_1}+\langle\Delta(h_{1,n}\hat Y_{23}^n)^2\rangle_{i_1}\}\\&
\geq P_1\sum\limits_{i_1}\eta_{i_1}\{\langle(\Delta\hat X_1^n)^2\rangle_{i_1}+\langle(\Delta\hat Y_1^n)^2\rangle_{i_1}\}
\geq P_1\langle\hat C_1^n\rangle
\end{split}
\end{equation}
\end{widetext}

Applying the same conditions as the above inequality on $S_{2,2}$ and $S_{3,3}$, we can get
\begin{equation}
\begin{split}
S_1^n+S_2^n+S_3^n&\geq P_1 S_{1,1}+P_2 S_{2,2}+P_3 S_{3,3}\\&\geq P_1 C_1^n+P_2 C_2^n+P_3 C_3^n\\&\geq min\{C_1^n, C_2^n, C_3^n\}
\label{a6}
\end{split}
\end{equation}
Violation of above inequality with any n is sufficient to confirm the genuine tripartite steering.

In our symmetric system, the Eq.(A8) can be simplified to
$3S_1^n\geq C_1^n$, and
\begin{equation}
\begin{split}
3S_1^n&=3(\langle(\Delta\hat U_{1,23}^n)^2\rangle+\langle(\Delta\hat V_{1,23}^n)^2\rangle)\\&=3\langle(\hat X_{1}^n)^2\rangle-3\langle\hat X_{1}^n\rangle^2+6g_{1,n}(\langle\hat X_{1}^n\hat X_{23}^n\rangle-\langle\hat X_{1}^n\rangle\langle\hat X_{23}^n\rangle)\\&+3g_{1,n}^2(\langle(\hat X_{23}^n)^2\rangle-\langle\hat X_{23}^n\rangle^2)+3\langle(\hat Y_{1}^n)^2\rangle-3\langle\hat Y_{1}^n\rangle^2\\&+6h_{1,n}(\langle\hat Y_{1}^n\hat Y_{23}^n\rangle-\langle\hat Y_{1}^n\rangle\langle\hat Y_{23}^n\rangle)\\&+3h_{1,n}^2(\langle(\hat Y_{23}^n)^2\rangle-\langle\hat Y_{23}^n\rangle^2),
\end{split}
\end{equation}
where the optimal gain parameters
$g_{1,n}=-h_{1,n}=\frac{-(\langle\hat X_{1}^n\hat X_{23}^n\rangle-\langle\hat X_{1}^n\rangle\langle\hat X_{23}^n\rangle)+(\langle\hat Y_{1}^n\hat Y_{23}^n\rangle-\langle\hat Y_{1}^n\rangle\langle\hat Y_{23}^n\rangle)}{(\langle(\hat X_{23}^n)^2\rangle-\langle\hat X_{23}^n\rangle^2)+(\langle(\hat Y_{23}^n)^2\rangle-\langle\hat Y_{23}^n\rangle^2)}$. Bringing the parameters back to $3S_1^n\geq C_1^n$, we can get
\begin{align}
&\big|\langle\hat a_1^n\hat a_2^n\hat a_3^n\rangle-\langle\hat a_1^n\rangle\langle\hat a_2^n\hat a_3^n\rangle\big|\leq
\nonumber\\
&~~~~~~\frac{1}{2}\sqrt{\big\langle\hat a_2^{\dag n}\hat a_2^n\hat a_3^{\dag n}\hat a_3^n\big\rangle+\big\langle\hat a_2^n\hat a_2^{\dag n}\hat a_3^n\hat a_3^{\dag n}\big\rangle-2\big\langle\hat a_2^n\hat a_3^n\big\rangle^2}\nonumber\\
&~~~~~~\times\sqrt{\big\langle\hat a_1^{\dag n}\hat a_1^n\big\rangle+\big\langle\hat a_1^n\hat a_1^{\dag n}\big\rangle-\frac{1}{6}C_{1}^n-2\big\langle\hat a_1^n\big\rangle^2}.
\end{align}
Violation of the above inequality can confirm the genuine tripartite steering.

When all the subsystems are constrained to be quantum states, we will get $\rho_{mix}'=P_1\sum\limits_{i_1}\eta_{i_1}\rho_{1}^{i_1}\rho_{23}^{i_1}+P_2\sum\limits_{i_2}\eta_{i_2}\rho_{2}^{i_2}\rho_{13}^{i_2}+P_3\sum\limits_{i_3}\eta_{i_3}\rho_{3}^{i_3}\rho_{12}^{i_3}$. Referring to the inequalities (A3), we can get
\begin{equation}
\begin{split}
E_1^n=\langle(\Delta\hat U_{1,23}^n)^2\rangle+\langle(\Delta\hat V_{1,23}^n)^2\rangle\geq C_1^n+|g_{1,n}h_{1,n}|C_{23}^n,\\
E_2^n=\langle(\Delta\hat U_{2,13}^n)^2\rangle+\langle(\Delta\hat V_{2,13}^n)^2\rangle\geq C_2^n+|g_{2,n}h_{2,n}|C_{13}^n,\\
E_3^n=\langle(\Delta\hat U_{3,12}^n)^2\rangle+\langle(\Delta\hat V_{3,12}^n)^2\rangle\geq C_3^n+|g_{3,n}h_{3,n}|C_{12}^n.
\end{split}
\end{equation}
Violation of the three inequalities above can confirm fully inseparable tripartite entanglement.

The same as the fully inseparable tripartite steering,
\begin{equation}
\begin{split}
E_1^n&=\langle(\Delta\hat U_{1,23}^n)^2\rangle+\langle(\Delta\hat V_{1,23}^n)^2\rangle\\&=\langle(\hat X_{1}^n)^2\rangle-\langle\hat X_{1}^n\rangle^2+2g_{1,n}(\langle\hat X_{1}^n\hat X_{23}^n\rangle-\langle\hat X_{1}^n\rangle\langle\hat X_{23}^n\rangle)\\&+g_{1,n}^2(\langle(\hat X_{23}^n)^2\rangle-\langle\hat X_{23}^n\rangle^2)
+\langle(\hat Y_{1}^n)^2\rangle-\langle\hat Y_{1}^n\rangle^2\\&+2h_{1,n}(\langle\hat Y_{1}^n\hat Y_{23}^n\rangle-\langle\hat Y_{1}^n\rangle\langle\hat Y_{23}^n\rangle)\\&+h_{1,n}^2(\langle(\hat Y_{23}^n)^2\rangle-\langle\hat Y_{23}^n\rangle^2)
\end{split}
\end{equation}
in our system, where the
$g_{1,n}=-h_{1,n}=\frac{-(\langle\hat X_{1}^n\hat X_{23}^n\rangle-\langle\hat X_{1}^n\rangle\langle\hat X_{23}^n\rangle)+(\langle\hat Y_{1}^n\hat Y_{23}^n\rangle-\langle\hat Y_{1}^n\rangle\langle\hat Y_{23}^n\rangle)}{(\langle(\hat X_{23}^n)^2\rangle-\langle\hat X_{23}^n\rangle^2)+(\langle(\hat Y_{23}^n)^2\rangle-\langle\hat Y_{23}^n\rangle^2)-C_{23}^n}$. Bringing the parameters back to the first formula in Eq.(A11), we can get a simplified inequality
\begin{align}
&\big|\langle\hat a_1^n\hat a_2^n\hat a_3^n\rangle-\langle\hat a_1^n\rangle\langle\hat a_2^n\hat a_3^n\rangle\big|\leq\nonumber\\
&~~~~~~~~~~~~~~~~~~
\sqrt{\big\langle\hat a_2^{\dag n}\hat a_2^n\hat a_3^{\dag n}\hat a_3^n\big\rangle-2\big\langle\hat a_2^n\hat a_3^n\big\rangle^2}\nonumber\\
&~~~~~~~~~~~~~~~~~~\times\sqrt{\big\langle\hat a_1^{\dag n}\hat a_1^n\big\rangle-\big\langle\hat a_1^n\big\rangle^2}.
\end{align}
Violation of the inequality above can confirm fully inseparable tripartite entanglement.

According to inequalities (A11) and $\rho_{mix}'$, we have
\begin{equation}
\begin{split}
E_1^n\geq P_{1}E_{1,1}^n+P_{2}E_{1,2}^n+P_{3}E_{1,3}^n\\
E_2^n\geq P_{1}E_{2,1}^n+P_{2}E_{2,2}^n+P_{3}E_{2,3}^n,\\
E_3^n\geq P_{1}E_{3,1}^n+P_{2}E_{3,2}^n+P_{3}E_{3,3}^n,
\end{split}
\end{equation}
where,
\begin{equation}
\begin{split}
E_{1,1}^n\geq P_1(\langle\hat C_1^n\rangle_1+|g_{1,n}h_{1,n}|\langle\hat C_{23}^n\rangle_1)
\end{split}
\end{equation}

\begin{widetext}
\begin{equation}
\begin{split}
&E_{1,2}^n=\langle(\Delta\hat U_{1,23}^n)^2\rangle_2+\langle(\Delta\hat V_{1,23}^n)^2\rangle_2\\&
\geq P_2\sum\limits_{i_2}\eta_{i_2}\{\langle(\hat X_1^n)^2+(g_{1,n}\hat X_{23}^n)^2+(\hat Y_1^n)^2+(h_{1,n}\hat Y_{23}^n)^2\rangle_{i_2}+2g_{1,n}\langle\hat X_1^n\hat X_{23}^n\rangle_{i_2}+2h_{1,n}\langle\hat Y_1^n\hat Y_{23}^n\rangle_{i_2}\}-P_2^2\langle\hat U_{1,23}^n\rangle_2^2-P_2^2\langle\hat V_{1,23}^n\rangle_2^2\\&
\geq P_2\sum\limits_{i_2}\eta_{i_2}\{\langle(\Delta\hat X_1^n)^2+(\Delta\hat Y_1^n)^2+(g_{1,n}\Delta\hat X_{23}^n)^2+(h_{1,n}\Delta\hat Y_{23}^n)^2\rangle_{i_2}+2g_{1,n}\langle\hat X_1^n\hat X_{23}^n\rangle_{i_2}+2h_{1,n}\langle\hat Y_1^n\hat Y_{23}^n\rangle_{i_2}\}\\&+(P_2\sum\limits_{i_2}\eta_{i_2}\langle\hat X_1^n\rangle_{i_2}^2-P_2^2\langle\hat X_1^n\rangle_2^2+P_2\sum\limits_{i_2}\eta_{i_2}\langle\hat Y_1^n\rangle_{i_2}^2-P_2^2\langle\hat Y_1^n\rangle_2^2)+(P_2\sum\limits_{i_2}\eta_{i_2}\langle\hat X_{23}^n\rangle_{i_2}^2-P_2^2\langle\hat X_{23}^n\rangle_2^2\\&+P_2\sum\limits_{i_2}\eta_{i_2}\langle\hat Y_{23}^n\rangle_{i_2}^2-P_2^2\langle\hat Y_{23}^n\rangle_2^2)-2P_2^2 g_{1,n}\langle\hat X_1^n\rangle_2\langle\hat X_{23}^n\rangle_2-2P_2^2 h_{1,n}\langle\hat Y_1^n\rangle_2\langle\hat Y_{23}^n\rangle_2
\\&\geq P_2(\langle\hat C_1^n\rangle_2+|g_{1,n}h_{1,n}|\langle\hat C_{23}^n\rangle_2+2g_{1,n}\langle\hat X_1^n\hat X_{23}^n\rangle_2+2h_{1,n}\langle\hat Y_1^n\hat Y_{23}^n\rangle_2)\\&-2P_2^2 g_{1,n}\langle\hat X_1^n\rangle_2\langle\hat X_{23}^n\rangle_2-2P_2^2 h_{1,n}\langle\hat Y_1^n\rangle_2\langle\hat Y_{23}^n\rangle_2
\end{split}
\end{equation}
\end{widetext}

Combining those inequalities in Eqs. (A14)-(A16), we find that
\begin{widetext}
\begin{equation}
\begin{split}
E_1^n+E_2^n+E_3^n&\geq C_1^n+|g_{1,n}h_{1,n}|C_{23}^n+C_2^n+|g_{2,n}h_{2,n}|C_{13}^n+C_3^n+|g_{3,n}h_{3,n}|C_{12}^n\\&+2(g_{2,n}+g_{3,n})P_1\langle\hat X_1^n\hat X_{23}^n\rangle_1+2(g_{1,n}+g_{3,n})P_2\langle\hat X_2^n\hat X_{13}^n\rangle_2+2(g_{1,n}+g_{2,n})P_3\langle\hat X_3^n\hat X_{12}^n\rangle_3\\&+2(h_{2,n}+h_{3,n})P_1\langle\hat Y_1^n\hat Y_{23}^n\rangle_1+2(h_{1,n}+h_{3,n})P_2\langle\hat Y_2^n\hat Y_{13}^n\rangle_2+2(h_{1,n}+h_{2,n})P_3\langle\hat Y_3^n\hat Y_{12}^n\rangle_3\\&-2P_2^2(g_{1,n}\langle\hat X_1^{n}\rangle_2\langle\hat X_{23}^{n}\rangle_2+h_{1,n}\langle\hat Y_1^{n}\rangle_2\langle\hat Y_{23}^{n}\rangle_2)-2P_3^2(g_{1,n}\langle\hat X_1^{n}\rangle_3\langle\hat X_{23}^{n}\rangle_3+h_{1,n}\langle\hat Y_1^{n}\rangle_3\langle\hat Y_{23}^{n}\rangle_3)\\&-2P_1^2(g_{2,n}\langle\hat X_2^{n}\rangle_1\langle\hat X_{13}^{n}\rangle_1+h_{2,n}\langle\hat Y_2^{n}\rangle_1\langle\hat Y_{13}^{n}\rangle_1)-2P_3^2(g_{2,n}\langle\hat X_2^{n}\rangle_3\langle\hat X_{13}^{n}\rangle_3+h_{2,n}\langle\hat Y_2^{n}\rangle_3\langle\hat Y_{13}^{n}\rangle_3)\\&-2P_1^2(g_{3,n}\langle\hat X_3^{n}\rangle_1\langle\hat X_{12}^{n}\rangle_1+h_{3,n}\langle\hat Y_3^{n}\rangle_1\langle\hat Y_{12}^{n}\rangle_1)-2P_2^2(g_{3,n}\langle\hat X_3^{n}\rangle_2\langle\hat X_{12}^{n}\rangle_2+h_{3,n}\langle\hat Y_3^{n}\rangle_2\langle\hat Y_{12}^{n}\rangle_2)
\label{a11}
\end{split}
\end{equation}
\end{widetext}
 Violation of the above inequality
is the condition for genuine tripartite entanglement.

In our system, the Eq.(A17) can be further simplified to
\begin{equation}
\begin{split}
3E_1^n&\geq 3C_1^n+3|g_{1,n}h_{1,n}|C_{23}^n+4g_{1,n}\langle\hat X_1^n\hat X_{23}^n\rangle+4h_{1,n}\langle\hat Y_1^n\hat Y_{23}^n\rangle\\&-4g_{1,n}\langle\hat X_1^n\rangle\langle\hat X_{23}^n\rangle-4h_{1,n}\langle\hat Y_1^n\rangle\langle\hat Y_{23}^n\rangle,
\end{split}
\end{equation}
where we use the operator properties
\begin{equation}
\begin{split}
&\langle\hat X_1^n\hat X_{23}^n\rangle=
P_1\langle\hat X_1^n\hat X_{23}^n\rangle_1+P_2\langle\hat X_2^n\hat X_{13}^n\rangle_2+P_3\langle\hat X_3^n\hat X_{12}^n\rangle_3,
\\&\langle\hat Y_1^n\hat Y_{23}^n\rangle=
P_1\langle\hat Y_1^n\hat Y_{23}^n\rangle_1+P_2\langle\hat Y_2^n\hat Y_{13}^n\rangle_2+P_3\langle\hat Y_3^n\hat Y_{12}^n\rangle_3,
\end{split}
\end{equation}
and
\begin{equation}
\begin{split}
&-P_1^2\langle\hat X_1^{n}\rangle_1\langle\hat X_{23}^{n}\rangle_1-P_2^2\langle\hat X_1^{n}\rangle_2\langle\hat X_{23}^{n}\rangle_2-P_3^2\langle\hat X_1^{n}\rangle_3\langle\hat X_{23}^{n}\rangle_3\\&\geq -P_1\langle\hat X_1^{n}\rangle_1\langle\hat X_{23}^{n}\rangle_1-P_2\langle\hat X_1^{n}\rangle_2\langle\hat X_{23}^{n}\rangle_2-P_3\langle\hat X_1^{n}\rangle_3\langle\hat X_{23}^{n}\rangle_3\\&= -\langle\hat X_1^{n}\rangle\langle\hat X_{23}^{n}\rangle,
\\&-P_1^2\langle\hat Y_1^{n}\rangle_1\langle\hat Y_{23}^{n}\rangle_1-P_2^2\langle\hat Y_1^{n}\rangle_2\langle\hat Y_{23}^{n}\rangle_2-P_3^2\langle\hat Y_1^{n}\rangle_3\langle\hat Y_{23}^{n}\rangle_3\\&\geq -P_1\langle\hat Y_1^{n}\rangle_1\langle\hat Y_{23}^{n}\rangle_1-P_2\langle\hat Y_1^{n}\rangle_2\langle\hat Y_{23}^{n}\rangle_2-P_3\langle\hat Y_1^{n}\rangle_3\langle\hat Y_{23}^{n}\rangle_3\\&= -\langle\hat Y_1^{n}\rangle\langle\hat Y_{23}^{n}\rangle.
\end{split}
\end{equation}

Then the same as the genuine tripartite steering,
\begin{equation}
\begin{split}
3E_1^n&=3(\langle(\Delta\hat U_{1,23}^n)^2\rangle+\langle(\Delta\hat V_{1,23}^n)^2\rangle)\\&=3\langle(\hat X_{1}^n)^2\rangle-3\langle\hat X_{1}^n\rangle^2+6g_{1,n}(\langle\hat X_{1}^n\hat X_{23}^n\rangle-\langle\hat X_{1}^n\rangle\langle\hat X_{23}^n\rangle)\\&+3g_{1,n}^2(\langle(\hat X_{23}^n)^2\rangle-\langle\hat X_{23}^n\rangle^2)
+3\langle(\hat Y_{1}^n)^2\rangle-3\langle\hat Y_{1}^n\rangle^2\\&+6h_{1,n}(\langle\hat Y_{1}^n\hat Y_{23}^n\rangle-\langle\hat Y_{1}^n\rangle\langle\hat Y_{23}^n\rangle)\\&+3h_{1,n}^2(\langle(\hat Y_{23}^n)^2\rangle-\langle\hat Y_{23}^n\rangle^2)
\end{split}
\end{equation}
in Eq.(A18), where the optimal gain parameters
$g_{1,n}=-h_{1,n}=\frac{-(\langle\hat X_{1}^n\hat X_{23}^n\rangle-\langle\hat X_{1}^n\rangle\langle\hat X_{23}^n\rangle)+(\langle\hat Y_{1}^n\hat Y_{23}^n\rangle-\langle\hat Y_{1}^n\rangle\langle\hat Y_{23}^n\rangle)}{3(\langle(\hat X_{23}^n)^2\rangle-\langle\hat X_{23}^n\rangle^2)+3(\langle(\hat Y_{23}^n)^2\rangle-\langle\hat Y_{23}^n\rangle^2)-3C_{23}^n}$. Bringing the parameters back to Eq.(A18), we can get
\begin{align}
&\big|\langle\hat a_1^n\hat a_2^n\hat a_3^n\rangle-\langle\hat a_1^n\rangle\langle\hat a_2^n\hat a_3^n\rangle\big|\leq\nonumber\\&~~~~~~~~~~~~~~~~~~~
3\sqrt{\big\langle\hat a_2^{\dag n}\hat a_2^n\hat a_3^{\dag n}\hat a_3^n\big\rangle-2\big\langle\hat a_2^n\hat a_3^n\big\rangle^2}\nonumber\\&~~~~~~~~~~~~~~~~~~~\times\sqrt{\big\langle\hat a_1^{\dag n}\hat a_1^n\big\rangle-\big\langle\hat a_1^n\big\rangle^2}.
\end{align}
Violation of the above inequality is the condition for genuine tripartite entanglement in our system.

\end{document}